\documentclass[12pt]{article}
\usepackage{xcolor}
\usepackage[colorlinks]{hyperref}
\usepackage{amsmath,amsfonts,amsthm,amssymb}
\usepackage{bm}
\usepackage{epsfig}  
\usepackage{subcaption}
\def\ftoday{{\sl {Le \number\day \space\ifcase\month 
\or janvier\or f\'evrier\or mars\or avril\or mai
\or juin\or juillet\or ao\^ut\or septembre\or octobre
\or novembre \or d\'ecembre\fi\space \number\year}}}    
\def\ptoday{{\sl {\number\day \space de\space \ifcase\month 
\or janeiro\or fevereiro\or mar{\c c}o\or abril\or maio
\or junho\or julho\or agosto\or setembro\or outubro
\or novembro \or dezembro\fi\space de\space \number\year}}}    
\def\gtoday{{\sl {Den \number\day. \ifcase\month 
\or Januar\or Februar\or M\"arz\or April\or Mai
\or Juni\or Juli\or August\or September\or Oktober
\or November \or Dezember\fi\space \number\year}}}    
\def\today{{\sl {\ifcase\month
\or January\or February\or March\or April\or May
\or June\or July\or August\or September\or October
\or November \or December\fi \space\number\day,\space 
                                            \number\year}}}
\newcommand{\journal}[4]{{\em #1~}#2\,(#3)\,#4}

\newcommand{\ijtp}{\journal {Int. J. Theor. Phys.}}

\newcommand{\pr}{\journal {Phys. Rev.}}

\newcommand{\jmp}{\journal {J. Math. Phys.}}

\newcommand{\rmp}{\journal {Rev. Mod. Phys.}}

\newcommand{\zp}{\journal {Z. Phys.}}

\newcommand{\Foundation}{\journal {Found. Phys.}}
\newcommand{\jsp}{\journal {J. Stat. Phys.}}
 
\renewcommand{\a}{\alpha}
\renewcommand{\b}{\beta}
\newcommand{\g}{\gamma}           
\renewcommand{\d}{\delta}         \newcommand{\D}{\Delta}

\newcommand{\la}{\lambda}        
\newcommand{\m}{\mu}
\newcommand{\n}{\nu}
\newcommand{\om}{\omega}         
\newcommand{\p}{\psi}              
\newcommand{\s}{\sigma}           \renewcommand{\S}{\Sigma}
\newcommand{\f}{{\phi}}           \newcommand{\F}{{\Phi}}

\newcommand{\es}{\\[3mm]}

\newcommand{\sla}{\raise.15ex\hbox{$/$}\kern -.57em} 
\newcommand{\Sla}{\raise.15ex\hbox{$/$}\kern -.70em}

\newcommand{\lp}{\left(}\newcommand{\rp}{\right)}

\newcommand{\complex}{{\kern .1em {\raise .47ex
\hbox {$\scriptscriptstyle |$}}
    \kern -.4em {\rm C}}}
\newcommand{\real}{{{\rm I} \kern -.19em {\rm R}}}
\newcommand{\rational}{{\kern .1em {\raise .47ex
\hbox{$\scripscriptstyle |$}}
    \kern -.35em {\rm Q}}}
\renewcommand{\natural}{{\vrule height 1.6ex width
.05em depth 0ex \kern -.35em {\rm N}}}

\newcommand{\half}{\dfrac{1}{2}}
\newcommand{\pa}{\partial}
\newcommand{\pad}[2]{{\dfrac{\partial #1}{\partial #2}}}

\newcommand{\dpad}[2]{{\displaystyle{\dfrac{\partial #1}{\partial #2}}}}

\newcommand{\dint}{\displaystyle{\int}}
\newcommand{\eg}{{\em e.g.,\ }}
\newcommand{\Eg}{{\em E.g.,\ }}
\newcommand{\ie}{{{\em i.e.},\ }}
\newcommand{\Ie}{{\em I.e.,\ }}

\newcommand{\etc}{{\em etc.\ }}

\newcommand{\twiddle}{\lower.9ex\rlap{$\kern -.1em\scriptstyle\sim$}}

\newcommand{\vev}[1]{\left\langle {#1}\right\rangle}
\newcommand{\Nabla}{\bm{\nabla}}
\newcommand{\equ}[1]{(\ref{#1})}
\newcommand{\eq}{\begin{equation}}
\newcommand{\eqn}[1]{\label{#1}\end{equation}}
\newcommand{\eea}{\end{eqnarray}}
\newcommand{\eqa}{\begin{eqnarray}}
\newcommand{\eqan}[1]{\label{#1}\end{eqnarray}}
\newcommand{\ba}{\begin{array}}
\newcommand{\ea}{\end{array}}
\newcommand{\eqac}{\begin{equation}\begin{array}{rcl}}
\newcommand{\eqacn}[1]{\end{array}\label{#1}\end{equation}}

\makeatletter 
\@addtoreset{equation}{section}  
\makeatother 
 
\newcommand{\bz}{\begin{enumerate}}
\newcommand{\ez}{\end{enumerate}}

\newcommand{\bx}{{\mathbf{x}}}

\newcommand{\bv}{{\mathbf{v}}}

\newcommand{\dbb}{\mathrm{B}}
\newcommand{\tfl}{t_\mathrm{flight}}

\begin{document}

\title{De Broglie - Bohm Cycles. \\ 
{\Large Free  Relativistic One-Half Particles}
\\[4mm]
{\hfill \it\small To my beloved Izabel}}
\author{Olivier Piguet\footnote{Pra\c ca Graccho Cardoso, 
76/504, 45015-180 Aracaju, SE, Brazil,
E-mail:  opiguet@yahoo.com}}
\date{June 6, 2023}    

\maketitle

\begin{abstract}

In the de Broglie-Bohm quantum theory, particles describe trajectories
determined by the flux associated with their wave function. 
These trajectories are studied here for relativistic spin-one-half particles.
Based in explicit numerical calculations for the case of a massless particle in dimension three space-time, it is shown that if the wave function is an eigenfunction of the total angular momentum, the trajectories -- here called ``de Broglie-Bohm cycles'' --
begin as circles of slowly increasing radius until a transition time at which they tend to follow straight lines. Arrival times at some detector, as well as their probability distribution are calculated, too. The chosen energy and momentum  parameters are of the orders of magnitude met in graphene's physics. 

\end{abstract}



\section{Introduction}	

Since its beginning in the first decades of XX$^{\rm th}$ 
century~\cite{Planck}-\cite{Dirac} Quantum Mechanics and its extension in the form of Quantum Field Theory led to an accurate description of atomic and subatomic phenomena, confirmed in an extraordinarily precise way by countless experiments. However, there is no such  broad consensus about  its interpretation. Various ones are  present in the literature, such as the Copenhagen~\cite{Copenhagen-int}, 
the Many-Worlds~\cite{Manyworld-int}, the Relational~\cite{Relational} or the de Broglie-Bohm (dBB) one. We will deal here with the latter interpretation, first proposed by Louis de Broglie~\cite{deBroglie} as the  "Pilot Wave Theory", later on formulated by 
David Bohm~\cite{Bohm,Bohm-Hiley} as the "Ontological Interpretation of Quantum Theory'' and finally, critically defended by John Bell in a series of papers reproduced in \cite{Bell-book}. This interpretation of Quantum Mechanics differs essentially from the largely more widespread Copenhagen interpretation by taking particle trajectories as elements of reality, \ie a particle really follows a trajectory, the latter being defined by ``guiding conditions'' first proposed by de Broglie. A probabilistic interpretation is maintained, but now in the sense of classical statistical mechanics. The  trajectory followed by a particle is fully defined by giving boundary conditions such as, \eg the coordinates of its initial position. The probability distribution of the particle following a particular trajectory is then given by the value of the squared modulus of the wave function taken at the initial time and position. Since this statistical distribution is equal to the ``usual'' (Copenhagen) quantum probability distribution and in particular satisfies the same conservation conditions, mean values of observables
evolve identically in both interpretations of Quantum Mechanics. 

The first aim of the present paper is to introduce the reader 
to the dBB theory by treating simple physical examples. 
Since the main peculiarity of this interpretation is the 
factual existence of trajectories, some effort will be made 
in the study of these trajectories and their properties.

The dBB trajectories we will calculate are those of a 
relativistic\footnote{Only the theory of a single particle 
is  considered here. 
See ~\cite{Holland-book,Durr-etal-relativity,Tumulka} for a 
discussion of the $N$-particle relativistic case.\label{N-relat}}, 
massive or massless, spin one half particle in dimension 
3 space-time. Its 
quantum state will be supposed to be described by an eigenfunction 
of the total angular momentum $J$ defined relatively to some space 
point\footnote{See~\cite{Bressanini-Ponti} for the calculation of 
such states in the case of the non-relativistic free particle.}. 
Stationary as well as packets of stationary wave functions will be considered. The main result for the latter is that the dBB trajectories consist of an initial phase of quasi circles whose  radius increases till a critical time at which the trajectory begins tending to a straight line -- the straight line expected for a classical particle with definite angular momentum. 

The concept of trajectory also allows for a natural definition of 
the arrival time at a detector, of a particle prepared in some 
initial state\cite{Daumer-etal,Durr-etal,Das-etal}\footnote{See 
~\cite{Durr-etal,Das-etal} for an interesting proposal for an experiment.}. 
Calculations of such arrival times  will be presented, together 
with their probability distributions.

The paper begins in Section \ref{dBB-theory} with an introduction 
to the dBB formalism and in Section \ref{r-particle} for its 
application to the  relativistic spin one half particle described by 
the Dirac equation. Numerical computations of dBB trajectories
and arrival times for electrons in the context of 
graphene's transport 
properties are presented in Section \ref{graphene}.
Final considerations are given in the Conclusion Section.

Most analytic and numerical calculations are made with the help of the
software Mathematica~\cite{Mathematica}. 

\section{Summary of the de Broglie-Bohm theory}\label{dBB-theory}

In the ''usual'' (\ie Copenhagen) interpretation of 
Quantum Mechanics~\cite{Copenhagen-int},
the dynamics of a physical system constituted of a single particle is described, in the Schr\"odinger picture, by a wave equation
\eq 
i \hbar \pad{\p (\bx,t)}{t} = \hat H \p (\bx,t),
\eqn{schrod-eq}
where $\hat{H}$ is a self-adjoint partial derivative operator acting
on a $N$-components wave function
\eq 
\p (\bx,t)=\lp\ba{c}\p_1(\bx,t)\\ \cdots\\ \p_N(\bx,t)\ea\rp,
\eqn{gen-spinor}
belonging to a Hilbert space, with scalar product and norm
 defined by\footnote{This holds \eg for a 
non-relativistic particle in general, or a spin 1/2 relativistic one.
We do not consider here cases such as the 
relativistic spin zero particle described by the 
Klein-Gordon equation.} 
\eq 
 \vev{\p |\F}= \dint_{\!\!\!
\mathbb{R}^d} d^d x\, \p ^\dagger(\bx,t) \f(\bx,t)
 = \dint_{\!\!\!
\mathbb{R}^d} d^d x \sum_{\a=1}^N (\p_\a^* \f_\a)(\bx,t),\qquad
||\p || =\vev{\p |\p }^{\frac12}.
\eqn{gen-norm}
$\bx=(x^i,i=1,\cdots,d)$ are the space coordinates, $d$ the space 
dimension and 
${}^\dagger$, ${}^*$ denote the hermitian and complex 
conjugate, respectively. 
The number of components, $N$, depends on the particle's spin 
and on the space dimension $d$. 
\Eg $N=1$ for a scalar (spin 0) particle, $N=2$ for a 
non-relativistic particle of spin 1/2 in any dimension, $N=2$ for 
a relativistic particle of spin 1/2 in 2 dimensions, 
$N=4$ for the same in 3 dimensions, \etc\cite{APais}.

The wave equation \equ{schrod-eq} implies the existence of a 
non-negative density
\eq
\rho(\bx,t)=\p^\dagger(\bx,t)\p(\bx,t),
\eqn{density-rho}
and an associate d-current density\footnote{Its explicit form 
will be given below for the cases studied there.} 
$\mathbf{j}(\bx,t)$ obeying a continuity equation
\eq  
\pa_t \rho +\Nabla \bm{j}=0,
\eqn{continuity}
which assures the constancy of the norm defined in \equ{gen-norm}.
Normalizing the norm to 1, one interprets $\rho$ as 
a probability density and $\bm{j}$ as a probability flux. 

In the Copenhagen theory, the state of the system is completely
characterized by the wave function $\p $, solution of the wave equation
\equ{schrod-eq}, with an arbitrarily given initial wave function
$\p (\bx,0)=\p _0(\bx)$. 
The de Broglie-Bohm (dBB) theory completes the characterization 
of the state of the system (the particle, here) by postulating the 
existence of a trajectory, determined by the de Broglie
 ``guidance conditions''~\cite{deBroglie} for the particle's velocity
 \eq 
 \bm{v}(\bx,t)=\bm{j}(\bx,t)/\rho(\bx,t),
 \eqn{dBB-cond}
 the dBB trajectory $\bx_\dbb(\bx_0,t)$ being then
 a solution of the set of differential  of equations
\eq 
\dpad{\bx_\dbb(\bx_0,t)}{t}=\bm{v}(\bx_\dbb(\bx_0,t),t),
\eqn{traj-eq}
with an arbitrarily given initial position 
$\bx_\dbb(\bx_0,0)=\bx_0$. In other words, the possible trajectories 
are the
integral lines of the dBB vector field $\bm{v}(\bx,t)$, labelled by their initial position $\bx_0$. 
Since the flux $\bm{j}$ and the density $\rho$
turn out to be both'' bilinear in $\p$ and $\p^\dagger$ 
(see later on), the dBB trajectories 
do not depend on the ``intensity'' of the wave function,
but only on its ``form''.

Recall that, in the quantum probabilistic interpretation of Copenhagen,
the density $\rho$ represents the probability of experimentally 
finding the particle at a given place at a given time.
On the other hand, the dBB theory treats this density 
in a more ``classical statistical mechanics'' way: Trajectories 
``really happen'', and their
 probability distribution, which amounts
to  the probability distribution of their initial positions 
$\bx_0$, is given~\cite{Durr-Goldstein-Zanghi} 
by the  density $\rho(\bx_0,0)$. 
The latter represents our lack of knowledge of the 
precise initial position. Thanks to the continuity equation 
\equ{continuity}, the probability for the particle being inside 
a co-moving volume\footnote{\Ie a volume whose boundary points 
move along the dBB trajectories.} 
$V(t)$ is constant in time, which sustains this
statistical interpretation, called ``thermal equilibrium'' 
in the literature~\cite{Durr-Goldstein-Zanghi}.

Such a formulation is called a theory with ``hidden variables'', 
the hidden variables of the present one being, \eg the components 
of the initial position $\bx_0$.

A heuristic justification for the guiding condition \equ{dBB-cond} 
may be found in analogy with fluid mechanics, considering 
$\rho$ and $\bm{v}$ as the fluid mass 
density and velocity field, respectively. 
With $\bm{j}=\rho \bm{v}$, Eq. \equ{continuity} 
has the form of the continuity 
equation  expressing the conservation of the fluid mass.

To summarize, the dBB theory proposes the existence of:
\begin{enumerate}
\item
A wave function or "guidance field"  \equ{gen-spinor}, solution of 
the Schrödinger-like wave equation \equ{schrod-eq}.
\item
A statistical ensemble of particle's trajectories 
$\bx_\mathrm{B}(\bx_0,t)$ 
as the integral lines of the dBB vector field \equ{dBB-cond},
solutions of the differential equations \equ{traj-eq}  parametrized
by the value of the initial position.
\end{enumerate}

All considerations made in this subsection generalize easily to
systems of $N$ particles, $\bx$ denoting a point of the 
$d\times N$  dimensional configuration space\footnote{At least
in the non-relativistic case. See footnote \ref{N-relat}.}.

\section{Free relativistic spin one half particle 
in 3-dim\-en\-sion\-al space-time}\label{r-particle}

\subsection{Dirac equation}

A free, spin 1/2
relativistic particle of mass $m$ in 
3-dimensional space-time is described in the usual theory by a 2-components 
spinor wave function
\eq
\p (x)=\lp\ba{cc}\p_1(x)\\ \p_2(x)\ea\rp,
\eqn{nr-spinor}
solution of the free Dirac equation
\eq
i\hbar c\g^\m \partial_\m\p (x) - m c^2\p (x) = 0,
\eqn{Dirac-eq}
where $x=(x^\m,\m=0,1,2)$ are the space-time coordinates space-time, whose 
 metric is $\eta_{\m\n}$ = $\mbox{diag}(1,-1,-1)$. The Dirac matrices 
 obey the anti-commutation rules
 $\{\g^\m,\g^\n\}=2\eta_{\m\n}$. Our choice for them is given in
 Appendix \ref{not-conv}.
$c$, $\hbar$ are the speed of light and the reduced Planck constant,
respectively.
The Dirac equation may be cast in the 
form\footnote{In fact Dirac's original form, reduced to 
3 space-time dimensions.}
of \equ{schrod-eq}:
\eq 
i \hbar \pad{\p (\bx,t)}{t} = \hat H \p (\bx,t),
\eqn{NRDirac}
with
\eq 
\hat{H}=-i\hbar c\,\a^i\pa_i + m c^2 \,\b ,
\eqn{Ham-Dirac}
with   $\a^i=\g^0\g^i$ and $\b=\g^0$ (see Appendix \ref{not-conv}). 
The density and the flux obeying the continuity equation \equ{continuity}
are given by
\eq 
\rho=\p^\dagger\p,\quad j^i = c\,\p^\dagger\a^i\p.
\eqn{rho-j}
The theory is relativistic: $c\rho$ and $j^i$ are the 
time and space components
of the  space-time 3-vector $j^\m=\bar{\psi}\g^\m\psi$, and  
the continuity equation reads 
  $\pa_\m j^\m=0$.
  
Another relativistic object is the scalar density
\eq 
\s(\bx,t)=\frac12\bar{\p}\p = \frac12\p^\dagger\b\p.
\eqn{scalar-density}
 $j^\m$ and  $\s$ fulfil the identity
\eq 
j^\m j_\m = 4\s^2,
\eqn{j^2=sigma^2}
consequence of the Pauli matrices identity
\eq
\sum_{i=1}^3\s^i_{\a\b}\s^i_{\g\d}=2\d_{\a\d}\d_{\b\g}-\d_{\a\b}\d_{\g\d}.
\eqn{current-spin-id}
Let us go now to the dBB theory.
The identity \equ{j^2=sigma^2} allows one to define the 
time-like 3-velocity field'
\eq 
u^\m=\dfrac{j^\m}{2| \s |},  \quad     u^\m u_\m=1,  \quad   u^0>0.
\eqn{3-velocity}
The relativistic form of the dBB guidance equation \equ{traj-eq} then reads\footnote{The present discussion is the reduction 
to 3-dimensional space-time of the one made 
by~\cite{Holland,Holland-book} in 4 dimensions.}
\eq 
\frac{dx^\m(\la)}{d\la}=u^\m(x(\la)),
\eqn{Dirac-traj-eq} 
with $\la$ as curve's parameter.
This equation is equivalent to \equ{traj-eq} and defines the space-time 
trajectories of the particle. The more practical non-covariant formalism based on Eqs. \equ{traj-eq} and \equ{NRDirac} will be used in the following.

\subsection{Eigenstates of the angular momentum and of the energy}\label{ang-energy}

We will look for the solutions of the Dirac equation which are eigenfunctions of
the total angular momentum and Hamiltonian operators. 
It will be useful to work in polar coordinates $r,\ \f$:
\[
x=r\cos\f,\quad y=r\sin\f.
\]
In these coordinates, the Hamiltonian operator \equ{Ham-Dirac} reads
\eq 
\hat{H} = -i\hbar c\lp \lp\a^1\cos\f+\a^2\sin\f\rp\pa_r
+\frac1r \lp\a^2\cos\f-\a^1\sin\f\rp\pa_\f\rp  + mc^2\b.
\eqn{Ham-Dirac-pol}
One easily checks, using the algebra of the Pauli matrices, 
that the total angular momentum operator with respect to the origin, which has a single 
component in the two-dimensional space's case,
\eq 
\hat{J} = \hat{L}+\hat{S},\quad\quad\quad\mbox{with}\
\quad \hat{L}=-i\hbar\pa_\f,\quad \hat{S}= \frac\hbar2\b,
\eqn{j-mom}
commutes with the Hamiltonian operator.
Spinor eigenfunctions of $\hat{J}$ with eigenvalue $j$
are readily found to be of the form
\eq 
\p(r,\f,t)=\lp\ba{l} {\rm e}^{i(j-\frac12)\f}f_1(r,t),\\
{\rm e}^{i(j+\frac12)\f}f_2(r,t)
\ea\rp.
\eqn{j-eigenfunction}
One directly sees that the requirement  
of the wave function to be single-valued requires $j$ to be half-integer. 
With the result \equ{j-eigenfunction}, solving the Dirac equation
\equ{Dirac-eq} amounts to solving the 
two radial equations for the functions $f_1(r,t)$ and $f_2(r,t)$:
\eq\ba{l}
i\hbar\lp r\lp \pa_t f_1 + c\,\pa_r f_2\rp + c(j+\half)\,f_2 \rp
-mc^2 r\, f_1\ = 0,\es
i\hbar\lp r\lp \pa_t f_2 + c\,\pa_r f_1\rp - c(j-\half)\,f_1 \rp
+mc^2 r\, f_2 = 0.
\ea\eqn{eq-rad-Dirac}
In terms of the radial functions $f_\a$, the probability density $\rho$ and 
the flux $\mathbf{j}$ read:
\eq\ba{l}
\rho(r,t) = |f_1(r,t)|^2+|f_2(r,t)|^2,\es
j_x(r,\f,t) = c\lp e^{i\f}f_1^*(r,t)f_2(r,t) 
+ e^{-i\f}f_2^*(r,t)f_1(r,t)\rp, \es
j_y(r,\f,t) =i c\lp -e^{i\f}f_1^*(r,t)f_2(r,t) 
+ e^{-i\f}f_2^*(r,t)f_1(r,t)\rp.
\ea\eqn{rho-j-Dirac}
The density $\rho$, as well as 
the radial and azimuthal components of
the flux $\mathbf{j}$,
\eq\ba{l}
j_r(r,t)=\cos\f\, j_x(r,\f,t)+\sin\f\, j_y(r,\f,t) = 
c\lp f_1^*(r,t)f_2(r,t) + f_2^*(r,t)f_1(r,t)\rp, \es
j_\f(r,t)=-\sin\f\, j_x(r,\f,t)+\cos\f\, j_y(r,\f,t) = 
ic\lp -f_1^*(r,t)f_2(r,t) + f_2^*(r,t)f_1(r,t)\rp,
\ea\eqn{j-polar-coor}
turn out to be independent of the angular coordinate.

\subsubsection{Stationary solutions:}\label{est-sol}

Since the Hamiltonian \equ{Ham-Dirac-pol}
and the total angular momentum \equ{j-mom} commute, we can
impose the stationarity condition 
\eq 
\hat{H}\p=\hbar\om_p\,\p.
\eqn{stationarity}
Thus the radial wave functions $f_\a(r,t)$ take the form
\eq 
f^{p,\,\rm stat}_\a(r,t) = 
e^{-i\om_p t}\, h^{p,\,\rm stat}_\a(r),\quad\a=1,2,
\eqn{stat-cond} 
$\hbar\om_p$ being the energy of the stationary state.
This leads to a pair of equations for the function
$h^{p,\,\rm stat}_\a(r)$, derived from \equ{eq-rad-Dirac} 
by substituting $i\pa_t$ by $\om_p$. 
The general solution of these equations is a superposition of 
the Bessel functions  of the first 
and second kind, $J_{j\pm1/2}(pr/\hbar)$ and 
$Y_{j\pm1/2}(pr/\hbar)$, with $p$ a function of $\om_p$ 
defined as the positive solution for $p$ of
\eq
\om_p = \frac1\hbar\sqrt{m^2c^4+p^2c^2}.
\eqn{omega(p)}
Square integrability of the wave function at $r=0$ leads us
to discard the solutions involving $Y_{j\pm1/2}$ 
because of the latter's singularity at the origin 
(see \equ{as-origin} in Appendix \ref{app-bessel}). 
The regular solution thus is
\eq 
h^{p,\,\rm stat}_1(r)= icp\,J_{j-1/2}(pr/\hbar),\quad
h^{p,\,\rm stat}_2(r) = -(\hbar\om_p-mc^2)J_{j+1/2}(pr/\hbar).
\eqn{stat-sol}
The general stationary solution of the Dirac
equation for 
angular momentum eigenstates, non-singular at the origin, then reads
\eq\ba{l}
\p^{p,\,\rm stat}(r,\f,t) 
=\lp\ba{l} {e}^{i(j-\frac12)\f}f_1^{p,\,\rm stat}(r,t),\es
e^{i(j+\frac12)\f}f_2^{p,\,\rm stat}(r,t)\ea\rp \\[10mm]
\phantom{\p^{p,\,\rm stat}(r,\f,t) }
= e^{-i\om_p t} \lp\ba{c} 
i c p\,{\rm e}^{i(j-1/2)\f}J_{j-1/2}(pr/\hbar),\es
-(\hbar\om_p-mc^2){\rm e}^{i(j+1/2)\f}J_{j+1/2}(pr/\hbar)
\ea\rp,
\ea\eqn{gen-st-sol}
with $\om_p$ given by \equ{omega(p)}. These spinors form a basis for the solutions of the Dirac equation, however an improper one since they are not square integrable due to the 
asymptotic behaviour of the Bessel functions shown in \equ{as-infinity}
of Appendix \ref{app-bessel}. 

We can nevertheless apply the 
dBB guidance principle, expressed in Eqs, \equ{dBB-cond} and
\equ{traj-eq}, to such a basis element. 
From the result \equ{gen-st-sol} we can compute explicitly the 
density $\rho$ given in \equ{rho-j-Dirac}: 
\eq 
\rho^{p,\,\rm stat}(r) =  c^2 p^2 J^2_{j-1/2}(pr/\hbar)
+ (\hbar\om_p-mc^2)^2 J^2_{j+1/2}(pr/\hbar),
\eqn{rho-Dirac-3D}
as well as the flux components \equ{j-polar-coor} which, divided through
$\rho$ according to the dBB condition, yields the radial and azimuthal
components of the velocity vector field:
\eq\ba{l}
v^{p,\,\rm stat}_r(r) = 0\,\es
v^{p,\,\rm stat}_\f(r) 
= 2c\dfrac{cp(\hbar\om_p-mc^2) J_{j-1/2}(pr/\hbar)J_{j+1/2}(pr/\hbar)}
{(cp)^2 J^2_{j-1/2}(pr/\hbar)
+ (\hbar\om_p-mc^2)^2J^2_{j+1/2}(pr/\hbar)}.
\ea\eqn{j-Dirac-3D}
Obviously all these expressions are time independent due 
to the stationarity condition. 
One sees that  the radial component of the velocity 
field is vanishing and that its
azimuthal component does not depend on the polar angle. 
Thus the dBB trajectories of the particle, defined as 
the integral curves of the velocity vector field, are circles 
of radius $r$ 
centred at the origin travelled at a constant velocity $v_\f$ 
and whose 
radius dependent value is bounded by $c$, 
in the massive as well as in the massless case. In all cases
the bound $c$ is effectively reached, for a discrete set of values of the radius. Fig.~\ref{Fig-rho-vphi-stat} in Subsection 
\ref{graphene} 
shows a typical behaviour of the 
azimuthal velocity  $v^{p,\,\rm stat}_\f(r)$ in the massless 
particle's case. 

One may observe that, in the massless case, in which $\hbar\om_p=pc$, 
changing the sign of the total angular momentum implies a change 
of sign of the velocity field: $v^{p,\,\rm stat}_\f(r)$ $\to$
$-v^{p,\,\rm stat}_\f(r)$, hence of the orientation of the 
trajectories. However this does not hold for the massive case, and 
also not for the more general wave packets examined in Subsection 
\equ{Gaussian-wave-packet}.

\subsubsection{Gaussian wave packet}\label{Gaussian-wave-packet}

 The general solution of the Dirac equation for 
angular momentum eigenstates, non-singular at the origin, reads
\eq 
\p(\bx,t)=  \dint_{\!\!\!0}^\infty\!\!\!\!\! dp\, a(p) \p^{p,\,\rm stat}(\bx,t),
\eqn{gen-sol-Dirac}
where $\p^{p,\,\rm stat}$ is the stationary solution 
\equ{gen-st-sol}
of energy\footnote{We restrict to 
positive energy contributions.}  $\hbar\om_p$ given by
\equ{omega(p)}, 
and the amplitude $a(p)$ is an arbitrary complex function, 
but constrained by the requirement 
of $\p(\bx,t)$ to be a square integrable function of $\bx$.

The angular momentum 
eigenstates we will consider in the following are described by the 
spinor wave functions of the form \equ{gen-sol-Dirac}, with $a(p)$
the Gaussian amplitude
\eq
a(p) = \sqrt{p}\, e^{-\dfrac{(p - p_0)^2}{2 \S^2}}.
\eqn{Gauss-ampl}
Here and in the rest of this paper we consider only positive energy 
solutions of the Dirac equations. 
\subsubsection{Mean values}\label{meanvalues}

We recall that mean values of observables obtained from the dBB or
from the Copenhagen theory coincide.

The wave functions \equ{gen-sol-Dirac} are normalizable and we 
can calculate the square norm $||\p||^2$ in the following way:
\[
\Vert\p\Vert^2 = 
\dint_{\!\!\!0}^{2\pi}\!\!\!\!\!d\f \dint_{\!\!\!0}^{\infty}\!\!\!\!\!dr\, r\,\rho(r,t) ,
\]
with $\rho(r,t)$ 
the density \equ{rho-j}. From \equ{gen-sol-Dirac} we get
\[\ba{c}
\Vert\p\Vert^2 = 2\pi \dint_{\!\!\!0}^{\infty}\!\!\!\!\!dr\,r
\dint_{\!\!\!0}^{\infty}\!\!\!\!\!dp 
\dint_{\!\!\!0}^{\infty}\!\!\!\!\!dp'
\,a(p)a(p')\sum_{\a=1}^2 
f^{p,\,\rm stat}_\a(r,t)^*f^{p',\,\rm stat}_\a(r,t)\es
= 2\pi \dint_{\!\!\!0}^{\infty}\!\!\!\!\!dp 
\dint_{\!\!\!0}^{\infty}\!\!\!\!\!dp'\,a(p)a(p')e^{i(\om_p-\om_{p'})}
\dint_{\!\!\!0}^{\infty}\!\!\!\!\!dr\,r   \es
\lp (mc^2+\hbar\om_p)(mc^2+\hbar\om_{p'})
J_{j-1/2}(pr/\hbar)J_{j-1/2}(pr'/\hbar)  +\hbar^2pp'
J_{j+1/2}(pr/\hbar)J_{j+1/2}(pr'/\hbar)  \rp.
\ea\]
From \equ{Gauss-ampl} and the completeness 
identity~\cite{compl-Bessel}
 for the Bessel functions:
\[
\dint_{\!\!\!0}^\infty dr\,r\, J_l(kr)J_l(k'r) = \frac1k \d(k-k'),
\]
one gets
\eq 
\Vert\p\Vert^2 = 2\pi\hbar^2\dint_{\!\!\!0}^{\infty}\!\!\!\!\!dp\,
\dfrac{1}{p}a^2(p)\lp c^2p^2+(\hbar\om_p-mc^2)^2 \rp.
\eqn{norm2}
In the same way one establishes expressions for the mean energy:
\[\ba{l}
\vev{E} = \dint_{\!\!\!0}^{2\pi}d\f \dint_{\!\!\!0}^{\infty}dr\, r\, 
\p^\dagger(r,\f,t)\hat{H}\,\p(r,\f,t)\,/\,\Vert\p\Vert^2,
\ea\]
where $\hat{H}$ is the Hamiltonian operator \equ{Ham-Dirac},
and for the standard energy deviation 
\[
\D E = \sqrt{\vev{E^2}-\vev{E}^2}.
\]
The result is
\eq\ba{l}
\vev{E} = \dfrac{2\pi\hbar^2}{\Vert\p\Vert^2} 
\dint_{\!\!\!0}^{\infty}\!\!\!\!\!dp\,
\dfrac{1}{p}a^2(p)\lp c^2p^2+(\hbar\om_p-mc^2)^2 \rp
\hbar\om_p ,\es
\D E = \lp  \dfrac{2\pi\hbar^2}{\Vert\p\Vert^2}
\dint_{\!\!\!0}^{\infty}\!\!\!\!\!dp\,
\dfrac{1}{p}a^2(p)\lp c^2p^2+(\hbar\om_p-mc^2)^2 \rp
(\hbar\om_p)^2 - \vev{E}^2 \rp^\frac12.
\ea\eqn{mean-energy}
Finally, a computation of the mean value of the spin 
operator $\hat{S}$ defined by \equ{j-mom} yields
\eq 
\vev{S} = \dfrac{\pi\hbar^3}{\Vert\p\Vert^2}
\dint_{\!\!\!0}^{\infty}\!\!\!\!\!dp\,
\dfrac{1}{p}a^2(p)\lp -c^2p^2+(\hbar\om_p-mc^2)^2 \rp.
\eqn{mean-S}
Substituting $\Vert\p\Vert^2$ in the denominator by its expression
\equ{norm2}, one sees that this mean value obeys the  bounds
$-\hbar/2<\vev{S}<\hbar/2$. 

All these integrals can be computed analytically in the massless case  
for the amplitude given by \equ{Gauss-ampl}. One then gets, for $m=0$,
\[
\Vert\p\Vert^2 = \pi\hbar^2 c^2 \S^3 
\lp \sqrt{\pi}(1+2z^2)(1+\mathrm{erf}(z)) + 2z e^{-z^2} \rp,
\]
\eq\vev{E} =c p_0\dfrac{\sqrt{\pi}z(3+2z^2)\lp 1+\mathrm{erf}(z)\rp
+2(1+z^2)e^{-z^2} }
{z(\sqrt{\pi}(1+2z^2)(1+\mathrm{erf}(z)) + 2z e^{-z^2})},
\eqn{meanE-explicit}
\eq 
\D E =c p_0\sqrt{
\dfrac{\pi(3+4z^4)(1+\mathrm{erf}(z))^2
+ 8\sqrt{\pi}z(-1+z^2)(1+\mathrm{erf}(z))e^{-z^2}
+4(-2+z^2)e^{-2z^2}}
{2z^2(\pi(1+2z^2)^2(1+\mathrm{erf}(z))^2
+ 4\sqrt{\pi}z(1+2z^2)(1+\mathrm{erf}(z))e^{-z^2}
+4z^2e^{-2z^2})}
}\eqn{DeltaE-explicit}
where we have set
\eq 
z=\dfrac{p_0}{\S},
\eqn{def-z}
and erf$(z)$ is the error function~\cite{error-function}. 

Finally,
the mean spin is null in this massless case: 
\eq 
\vev{S}=0,
\eqn{mean-S-massless}
which implies
$\vev{L}=\hbar j$ for the orbital angular momentum since the 
states considered are eigenstates of the total angular 
momentum with eigenvalue $\hbar j$. Associated to this mean orbital angular momentum, we can define an {\it $L$-radius}
\eq 
r_L = \dfrac{\vev{L}}{\vev{p}} = \dfrac{c\,\hbar\, j}{\vev{E}},
\eqn{L-radius}
where $\vev{p}$ is the mean value of the momentum $p$, equal to
$\vev{E}/c$ in the present massless case. This definition mimics the 
classical relation between angular momentum and radius for a 
uniform circular motion. 

As one may expect, in the case of a very narrow width of the amplitude \equ{Gauss-ampl},
\ie $z\gg1$, the energy and  the energy uncertainty approximate to the values
\eq
\vev{E} \simeq cp_0,\quad \D E \simeq \dfrac{c}{\sqrt2}\S.
\eqn{E-DE-approx}
The results \equ{mean-energy} and \equ{mean-S} allow us to 
identify the expression 
\eq 
\tilde\rho(p) =  a^2(p)\dfrac{c^2p^2+(\hbar\om_p-mc^2)^2}{p^2}
\eqn{tilde-rho}
up to a due normalization, as the probability density in 
$p$-space, conjugate to the $x$-space probability defined in
\equ{rho-j}.

\subsection{{\em j} - electrons in graphene}\label{graphene}

Let us denote free electrons in eigenstates of the total angular 
momentum with eigenvalue $j$ as ``$j$-electrons''. 
Free electrons in mono-layer graphene~\cite{Katsnelson}
with energy less than 
$E_{\rm crit}\approx160$ meV obey approximatively a relativistic-like
massless dispersion law $E\approx p\,c$, where $p$ is the linear 
momentum and the ``velocity of light''\footnote{This is the 
so-called critical velocity, which we will denote by $c$.}
$c\approx 10^6\, \mbox{m\,s}{}^{-1}$.
The dynamics of these electron is that of free massless particles
in dimension three space-time with pseudo-spin 1/2 obeying
the Dirac equation \equ{Dirac-eq} with $m=0$~\cite{Das-Sarma}.
Pseudo-spin is a chirality effect 
due to the peculiar crystalline structure of graphene and 
should not be confused with the usual spin. 
Nevertheless, since the wave function obeys the Dirac equation 
\equ{Dirac-eq}, the pseudo-spin adds itself to the orbital 
angular momentum yielding the conserved total angular momentum 
$j$ \equ{j-mom} as explained in 
Subsection \ref{ang-energy}.

We will consider first stationary states and then states defined by Gaussian-like wave packets.
Conventions and units used in the following are described in Appendix 
\ref{not-conv}.

\subsubsection{Stationary states in graphene}

As already shown in Subsection \ref{est-sol}, the dBB trajectories 
associated to the stationary wave functions \equ{gen-st-sol}
are 
circles centred at the origin, travelled at a constant 
dBB velocity $v^{p,\,\rm stat}_\f$ which
depends on the radius $r$ according to \equ{j-Dirac-3D}. For 
$m=0$, this velocity reads
\eq 
v^{p,\,\rm stat}_\f(r) 
= 2c\dfrac{ J_{j-1/2}(pr/\hbar)J_{j+1/2}(pr/\hbar)}
{J^2_{j-1/2}(pr/\hbar) + J^2_{j+1/2}(pr/\hbar)}.
\eqn{vpfi-m=0}
The dBB velocity value oscillates between $-c$ and $c$. 
\begin{figure}[!t]
\includegraphics[width=0.45\textwidth]{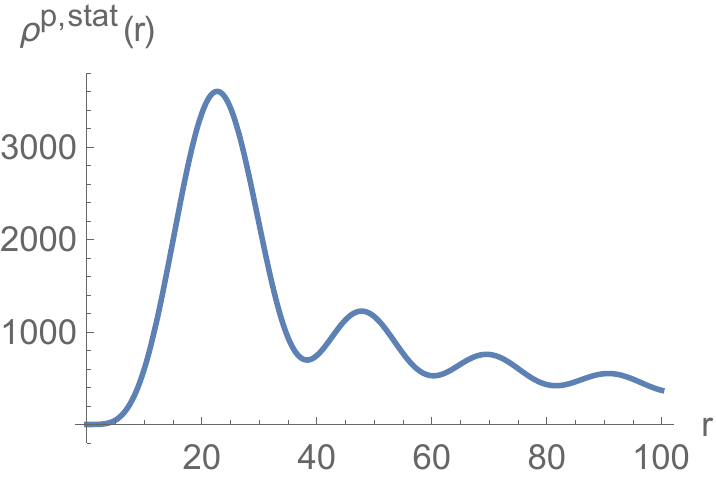}
\hspace{10mm}
\includegraphics[width=0.45\textwidth]{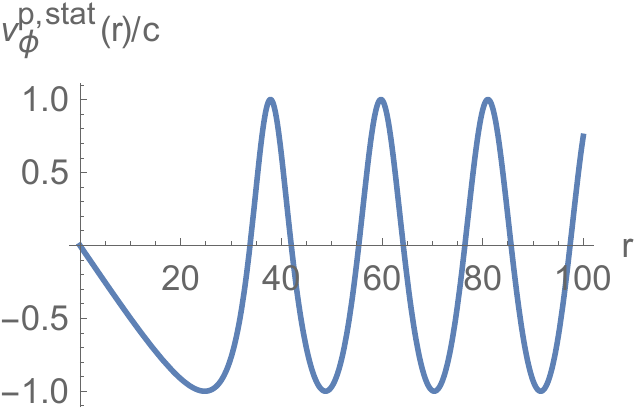}
\caption{Stationary case: density $\rho^{p,\,\rm stat}(r)$ 
and azimuthal velocity
$v_\f^{p,\,\rm stat}(r)/c$ for $j=5/2$, $p=10^{-4}$ meV/$c$
($E=100$ meV).} 
\label{Fig-rho-vphi-stat}
\end{figure}
A typical behaviour, as a function of the radius, of the probability density and of the dBB velocity, which depend on the energy $E=c\,p$
and on the angular momentum $j$, 
is shown in Fig.~\ref{Fig-rho-vphi-stat} for $j=5/2$ and $E=100$ meV.
The most probable radius $\hat{r}_j$, is given by the position  of 
the first maximum of the probability density 
$\rho^{p,\,\rm stat}$ \equ{rho-Dirac-3D} 
(See Fig. \ref{Fig-rho-vphi-stat}), 
which for $m=0$ reads
\eq 
\rho^{p,\,\rm stat}(r) =  c^2 p^2 \lp J^2_{j-1/2}(pr/\hbar)
+  J^2_{j+1/2}(pr/\hbar)\rp.
\eqn{rho-massless}
Since $r$ appears in the combination $pr/\hbar$, 
the most probable radius $\hat{r}_j$ may be written as a
function of $p$: 
\eq
\hat{r}_j(p)= \a_j\dfrac{\hbar}{p},
\eqn{most-prob-r}
the $j$-dependent coefficients $\a_j$ being shown in Table 
\ref{table-r-hat} for some values of $j$. 
\begin{table}[!tb]\begin{center}
\begin{tabular}{|c||c|c|c|c|c|c|c|c|c|}
\hline
$j$ & 1/2 & 3/2 & 5/2 & 7/2 & 9/2 & 11/2 &13/2 & 15/2 & 17/2 \\[1mm]
\hline
$\a_j$&  0& 2.19& 3.45& 4.61& 5.74& 
6.84& 7.93&  9.01& 10.08\\[1mm]
\hline
\end{tabular}
\caption{Coefficients $\a_j$ of Eq. \equ{most-prob-r} as a  
function of the angular momentum $j$.}
\label{table-r-hat}
\end{center}\end{table}
These results can be 
taken as an approximation for the more realistic case of a 
wave packet with a very small energy width.

\subsubsection{Gaussian wave packets in graphene}

\begin{figure}[p]
\begin{subfigure}[b]{0.3\textwidth}
\centering
\includegraphics[width=1.\textwidth]{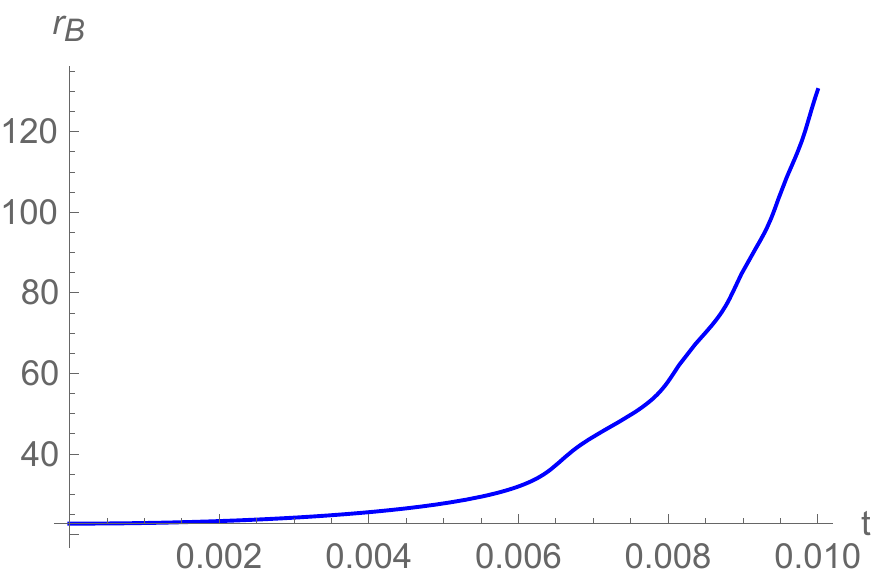}
\caption{} 
   \label{fig-gauss2a}
\end{subfigure}%
\begin{subfigure}[b]{0.3\textwidth}
\centering
\includegraphics[width=1.\textwidth]{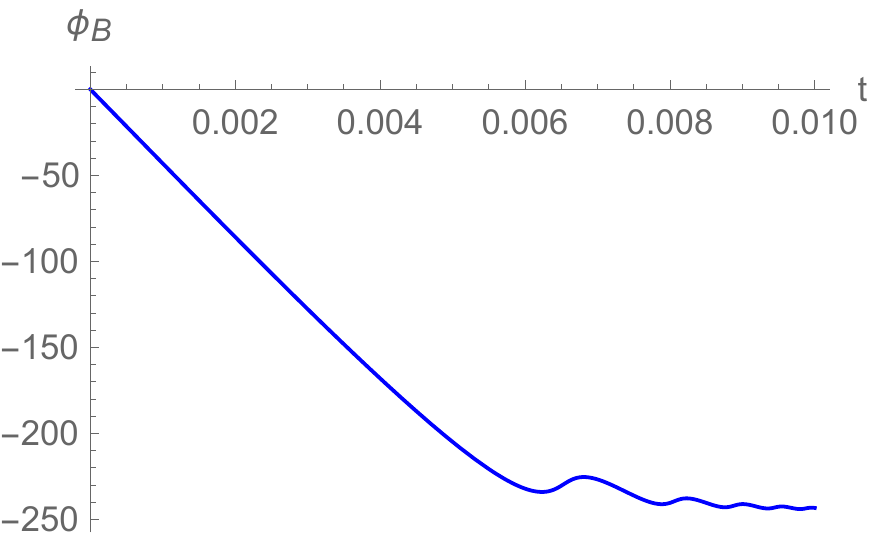}
\caption{} 
   \label{fig-gauss2b}
\end{subfigure}%
\begin{subfigure}[b]{0.4\textwidth}
\centering
\includegraphics[width=1.\textwidth]{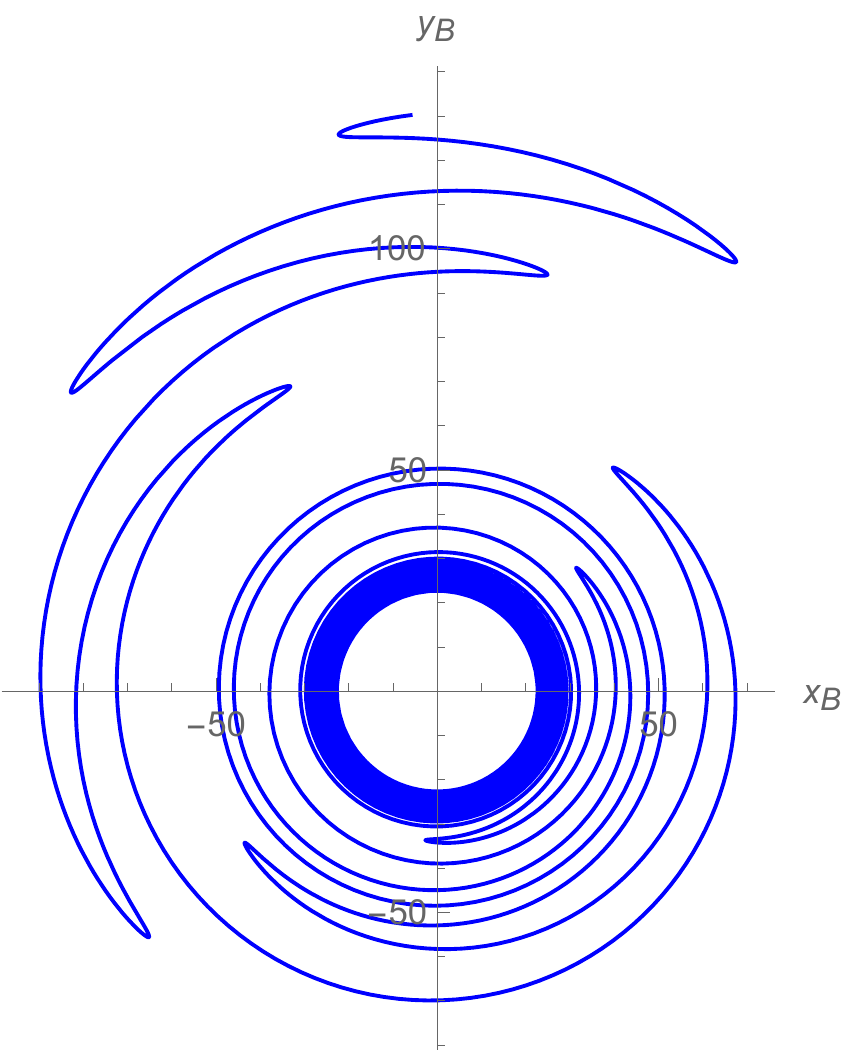}
\caption{} 
   \label{fig-gauss2c}
\end{subfigure}%
\caption{The most probable dBB tractory $\bx_\mathrm{B}(\hat{r}_0,t)$
for Gaussian wave packet parameters
 $j=5/2$, $p_0=10^{-4}$ meV/$c$ (Peak energy $E_0=100$ meV), 
and $\S=10^{-7}$ meV/$c$. This trajectory is fixed by the 
initial condition parameter $=\hat{r}_0=22.7$.
$\hat{r}_0$ is the position of the peak of
the probability density $\rho$ at $t=0$ (the blue point in Fig. 
\ref{fig-gauss4a}.)
\\
 (a) Radial coordinate $r_\dbb(\hat{r}_0,t)$.\\
 (b)  Azimuthal coordinate $\f_\dbb(\hat{r}_0t)$.\\
 (c) dBB trajectory in the $(x,y)$ plane for  $0\le t\le0.010$ ns.
 The trajectory performs 38 loops before the critical time 
 (decay time) $\tau\sim 0.006$ ns.}
\label{fig-gauss2}
\end{figure}
\begin{figure}[!ht]
\begin{subfigure}[b]{0.3\textwidth}
\centering
\includegraphics[width=1.\textwidth]{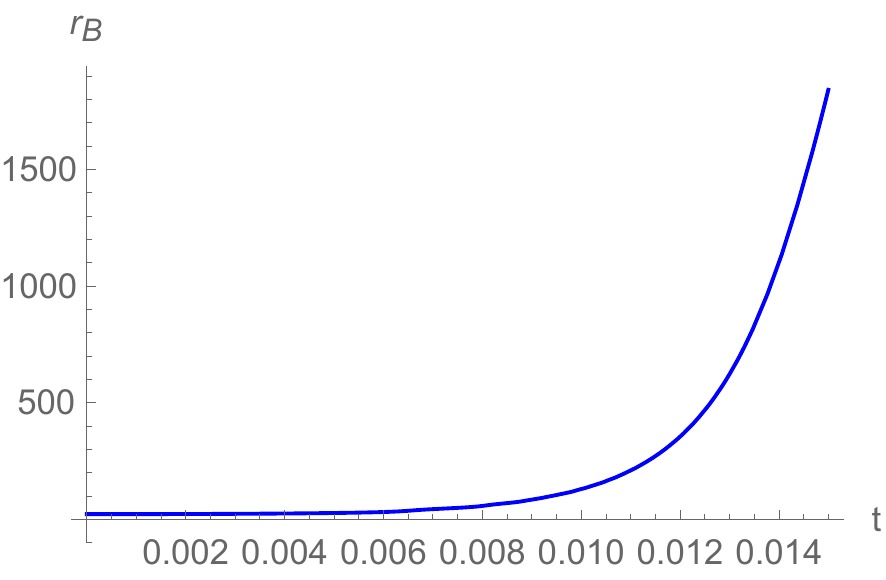}
\caption{} 
   \label{fig-gauss3a}
\end{subfigure}%
\begin{subfigure}[b]{0.3\textwidth}
\centering
\includegraphics[width=1.\textwidth]{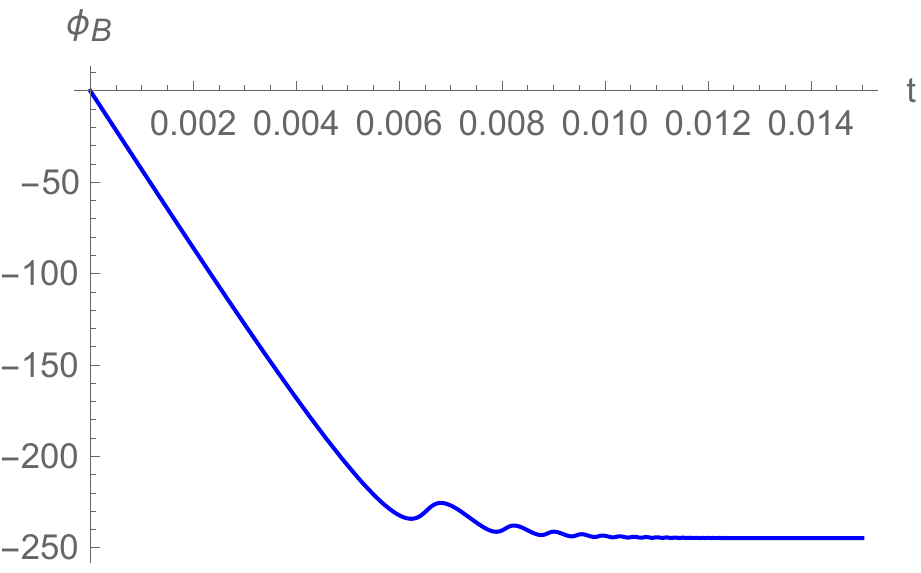}
\caption{} 
   \label{fig-gauss3b}
\end{subfigure}%
\begin{subfigure}[b]{0.4\textwidth}
\centering
\includegraphics[width=1.\textwidth]{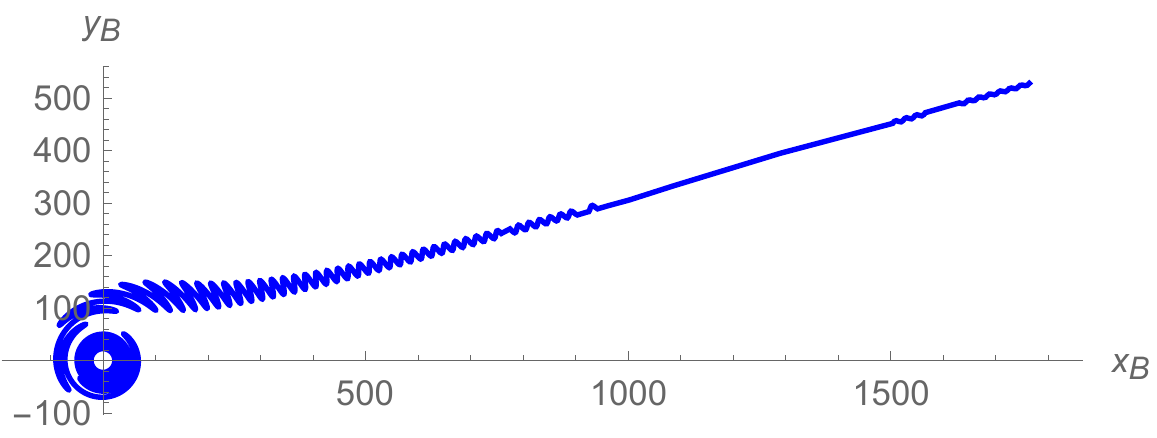}
\caption{} 
   \label{fig-gauss3c}
\end{subfigure}%
\caption{dBB trajectories for Gaussian wave packet: 
same parametrization as in Fig. 
\ref{fig-gauss2}, but with the larger time scale $0\le t\le0.015$ ns.}
\label{fig-gauss3}
\end{figure}

We turn now to the Gaussian wave functions defined by
\equ{gen-sol-Dirac} and \equ{Gauss-ampl}, which are normalizable. 
We shall denote by
\[
\bx_\mathrm{B}(r_0,t)=\lp r_\mathrm{B}(r_0,t),\ \f_\mathrm{B}(r_0,t)\rp,
\]
the  (polar) coordinates of the dBB trajectory solution of
the trajectory equation  
\equ{traj-eq}, fixed by the initial position 
\eq 
\bx_\mathrm{B}(r_0,0) = 
(r_\mathrm{B}(r_0,0),\f_\mathrm{B}(r_0,0) = (r_0,0).
\eqn{r0hat}
We have made explicit, in our notation, the dependence on 
the trajectory parameter $r_0$.

\begin{figure}[!hb]
\begin{subfigure}[b]{.4\textwidth}
\centering
\includegraphics[width=1.\textwidth]{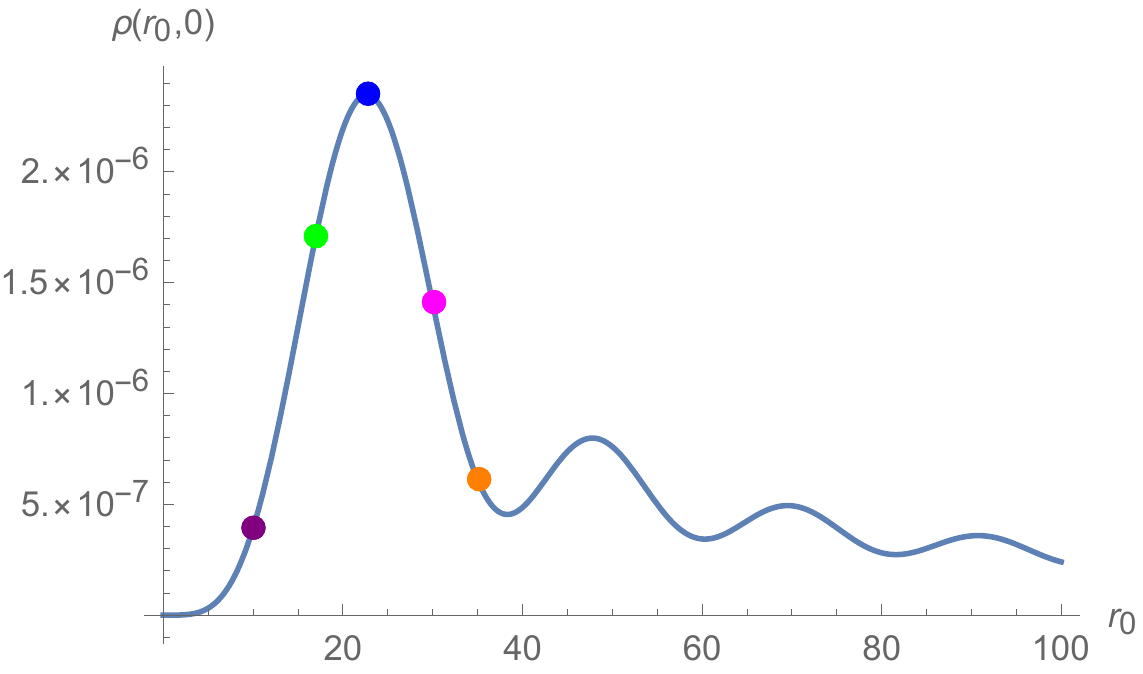}
\caption{} 
   \label{fig-gauss4a}
\end{subfigure}%

\begin{subfigure}[b]{0.3\textwidth}
\centering
\includegraphics[width=1.\textwidth]{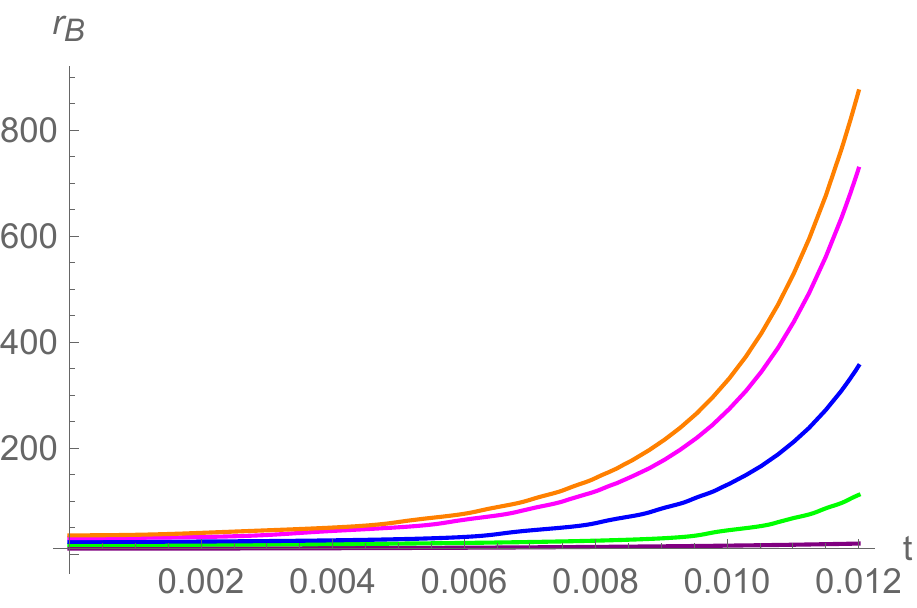}
\caption{} 
   \label{fig-gauss4b}
\end{subfigure}%
\begin{subfigure}[b]{0.3\textwidth}
\centering
\includegraphics[width=1.\textwidth]{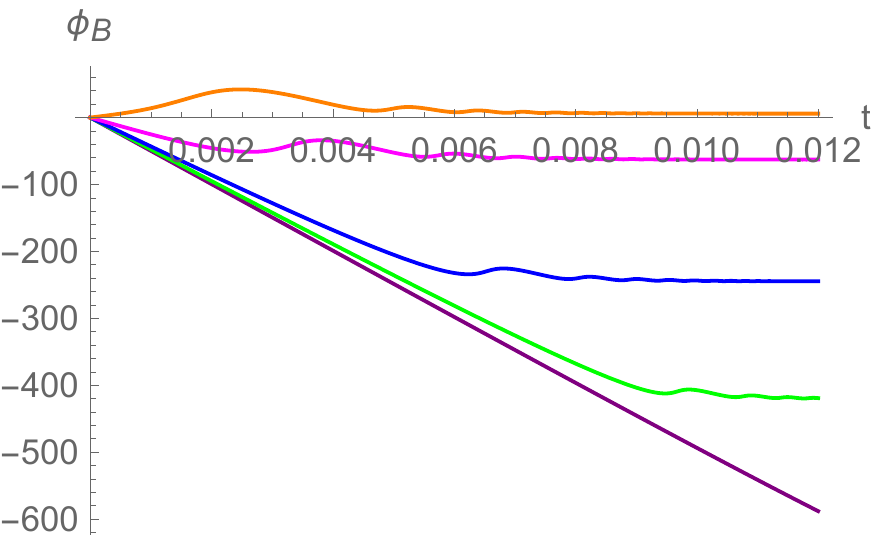}
\caption{} 
   \label{fig-gauss4c}
\end{subfigure}%
\begin{subfigure}[b]{0.4\textwidth}
\centering
\includegraphics[width=1.\textwidth]{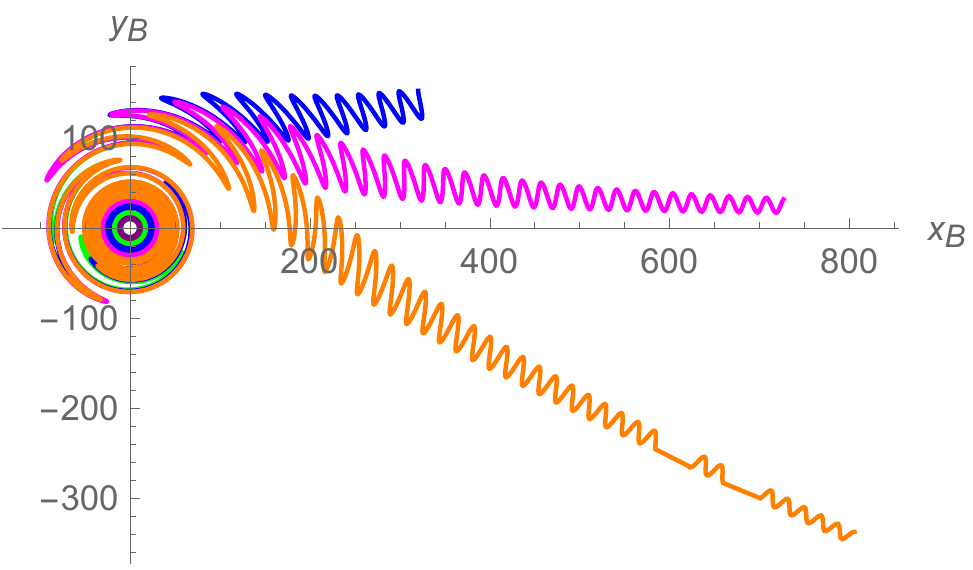}
\caption{} 
   \label{fig-gauss4d}
\end{subfigure}%
\caption{dBB trajectories for Gaussian wave packet: same parametrization
as in Figs. \ref{fig-gauss2} and \ref{fig-gauss3}, but with five 
trajectories corresponding to the initial radial positions $r_0$
= 10, 17, 22.7, 30 and 35 nm. The respective numbers of 
trajectory's closed loops are 93, 66, 38, 10 and 1.
Their respective relative probabilities are proportional to the 
heights of the coloured dots  
in  the sub-figure (a) showing the initial probability 
density $\rho(r_0,0)$ as a function of the initial radial position $r_0$.}
\label{fig-gauss4}
\end{figure}

Figs.  \ref{fig-gauss2} and 
\ref{fig-gauss3} show the most probable dBB trajectory 
$\bx_\mathrm{B}(\hat{r}_0,t)$ for a particular
wave function's  parametrization in which the 
energy dispersion is very small, \ie $\S\ll p_0$.
``Most probable'' means that the initial particle's 
position parameter $\hat{r}_0$ is the value of the 
radial coordinate $r$ which
maximizes the initial probability density $\rho(r,0)$. 
This value, equal to 22.7 nm in the present 
example\footnote{This value is very near of the one 
corresponding to  the stationary wave function with same 
$j=5/2$ and $p$ equal to the momentum parameter $p_0=10^{-4}$. 
Indeed, Eq. \equ{most-prob-r} together with Table \ref{table-r-hat} 
yield $\hat{r}$= 22.6.},
corresponds to the blue dot in Fig. \ref{fig-gauss4a}. 
Thus, the behaviour of this trajectory,
shown by Figs. \ref{fig-gauss2c} or \ref{fig-gauss3c}
in the $(x,y)$-plane, 
is very similar to the circular one shown in
the corresponding stationary solution, but only up to a certain 
critical ``decay time'' $\tau_\mathrm{obs}$, approximately equal to 
0.006 ns in this example. For later times the trajectory 
tends to a 
straight line, reproducing the expected classical behaviour. 
This is best observed in Figs. \ref{fig-gauss2b} or 
\ref{fig-gauss3b} which show
the time behaviour of the azimuthal angle $\f$: 
the angular velocity 
is almost constant until the time $\tau_\mathrm{obs}$, and almost 
vanishes thereafter. This decay time marks the transition from the
 almost circular regime to an almost straight-way, classical-like, regime.
 
 A theoretical lower bound for  the decay time  $\tau$ may  
be computed from the quantum
"time-energy uncertainty principle"~\cite{time-energy-uncertainty}:
\eq 
\tau \ge \tau_\mathrm{min} = \frac{\hbar}{2\,\D E},
\eqn{time-incert}
where $\D E$ is the quantum energy uncertainty given by 
\equ{mean-S}. 
In our example,  $\tau_\mathrm{min}$ = 0.00465 ns: this is
the order of magnitude of the 
observed decay time for the most probable trajectory, 
$\tau_\mathrm{obs}$ $\sim$  0.006 nm, and the 
uncertainty inequality \equ{time-incert} is obeyed.

A simple check of the quasi-classical nature of the trajectories for large times
consists in comparing the quantum mean orbital angular momentum  $\vev{L}=\hbar j$ 
given after Eq. \equ{mean-S-massless}, with the orbital momentum
 of a massless free classical particle of energy $E$ 
running along a straight trajectory $r(t)$ defined by its position and velocity 
at some time $t$, given by
\eq
L_{\rm class} = \dfrac{E}{c^2} r(t)v_\f(t),
\eqn{L-class}
where $v_\f(t)$ is its azimuthal velocity. Inserting for $E$ in 
the latter expression  the quantum mean energy $\vev{E}$ \equ{meanE-explicit}, 
and for $r(t)$, $v_\f(t)$ the dBB trajectory function $r_\dbb(t)$ and the
corresponding azimuthal velocity,
and equating it to the quantum expectation value $\vev{L}=\hbar j$,  
yields a value for $L_{\rm class}$ which, at a sufficiently large time, 
should tend to the value $\hbar j$  
characterizing the considered wave function. For the particular case 
shown in Figs. \ref{fig-gauss2} and \ref{fig-gauss3}, with $j=5/2$  
and a transition time 
$\tau_{\rm obs}\sim 0.006$ ns, we have checked that the resulting value
for $L_{\rm class}$ at the rather larger time $t=0.030$ ns, indeed coincides with  
$\hbar j$ up top an error of the order of $10^{-5}$. Such a concordance holds 
for the most probable or almost most probable trajectories, but tends to 
diminish for the less probable ones, \ie those for which the 
initial radial coordinate  lies farther of the probability distribution's peak 
position shown, \eg in Fig. \ref{fig-gauss4a}. 

Fig. \ref{fig-gauss4} shows trajectories, calculated for the same wave function as for  Figs. 
\ref{fig-gauss2} and \ref{fig-gauss3}, for
five initial radial positions.

\begin{figure}[!ht]
\centering
\includegraphics[width=0.5\textwidth]{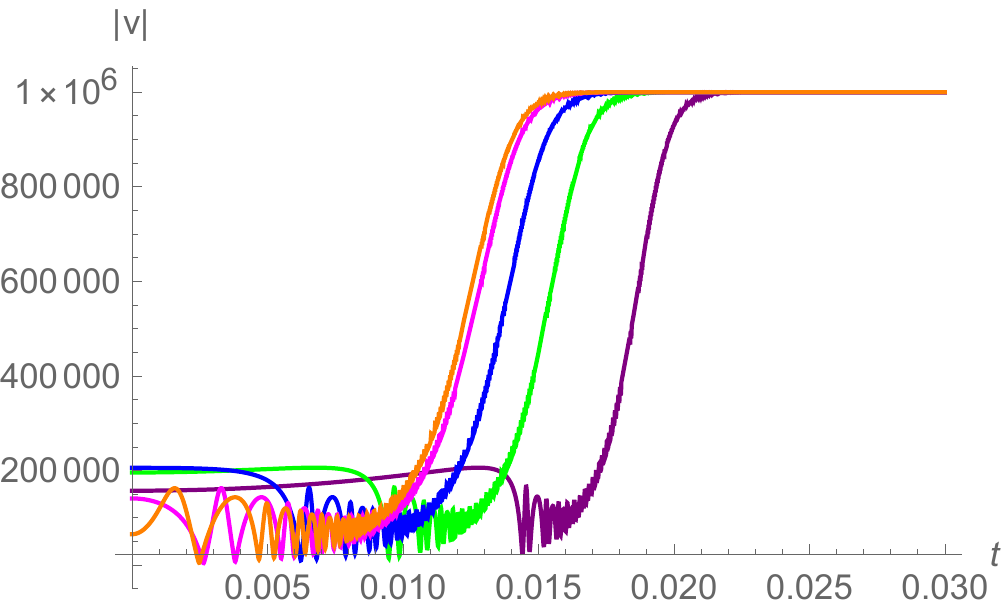}
   \label{fig-gauss5c}
\caption{Absolute velocity $|v|$  
as a function of $t$ for the solutions shown in 
Figs. \ref{fig-gauss2}, \ref{fig-gauss3} and \ref{fig-gauss4}. Colors
correspond to the five different trajectories exhibited in Fig. 
\ref{fig-gauss4}.}
\label{fig-gauss5}
\end{figure}

Fig. \ref{fig-gauss5} shows the time evolution of the 
 absolute velocity $|\bv|(r_0,t)$
for the state already exhibited in the figures above. $|\bv|$ is shown for five dBB trajectories,
characterized by their initial position parameters $r_0$ and colours as 
in Fig. \ref{fig-gauss4}. 
One  observes that,
for a each trajectory, the absolute velocity tends to the "velocity of light"'s 
value $c=10^6$ nm/ns above a certain threshold.
For example, for the blue trajectory, which corresponds to the initial 
radial position $r_0=22.7$, the threshold is at $t\sim0.017$. It is a bit higher than the decay time $\tau$ ($\sim0.006$ ns here).

\begin{table}[!tb]
\begin{center}
\begin{tabular}{|c|c||c|c|c|c|c|c|c||c|}
\hline
$p_0{}_\mathrm{(\frac{meV}{c})}$ &$\S{}_\mathrm{(\frac{meV}{c})}$ 
&$\vev{E}{}_\mathrm{(meV)}$ & $\D E{}_\mathrm{(meV)}$ 
&$\tau_\mathrm{min}{}_\mathrm{(ns)}$ 
&$\tau_\mathrm{obs}{}_\mathrm{(ns)}$ 
 &$r_L{}_\mathrm{(nm)}$ &$\hat{r}_0{}_\mathrm{(nm)}$ &$N_\mathrm{loops}$&$j$ \\[2mm]
\hline\hline
& & & & &0.3 &32.91 &0  &757 &1/2\\[1mm]
\cline{6-10}
&$10^{-8}$ &10.0000 &0.00707 &0.0465 &0.06 &164.6 &227 &39&5/2 \\[1mm]
\cline{6-10}
& & & & &0.05 &362.0 &450 &20&11/2 \\[1mm]
\cline{6-10}
$10^{-5}$
& & & & &0.05 &3258. &3450 &3 &99/2 \\[1mm]
\cline{2-10}
& & & & &0.001 &32.59 &0 &3 &1/2\\[1mm]
\cline{6-10}
&$10^{-6}$ &10.0995 &0.704 &0.000468 &-- &162.9 &220 &$<1$&5/2 \\[1mm]
\cline{6-10}
&  & & & &-- &358.4 &435 &$<1$&11/2\\[1mm]
\cline{6-10}
& & & & &-- &3226. &3250 &$<1$ &99/2 \\[1mm]
\hline
& & & & &0.027 &3.291 &0 &676 &1/2 \\[1mm]
\cline{6-10}
&$10^{-7}$ &100.000 &0.0707 &0.00465 &0.006 &16.46 &22.7 &39 &5/2 \\[1mm]
\cline{6-10}
& & & & &0.006 &36.20 &45.0 &20 &11/2 \\[1mm]
\cline{6-10}
$10^{-4}$& & & & &0.005 &325.8 &345 &3 &99/2 \\[1mm]
\cline{2-10}
& & & & &0.0001 &3.259 &0 &2 &1/2\\[1mm]
\cline{6-10}
&$10^{-5}$ &100.995 &7.04 &0.0000468 &-- &16.29 &22.0 &$<1$ &5/2\\[1mm]
\cline{6-10}
& & & & &-- &35.84 &43.5 &$<1$&11/2\\[1mm]
\cline{6-10}
& & & & &-- &322.6 &325 &$<1$ &99/2\\[1mm]
\hline
\end{tabular}
\caption{Mean energy $\vev{E}$, standard energy deviation $\D E$,
decay time lower bound $\tau_\mathrm{min}$, 
observed decay time  $\tau_\mathrm{obs}$,
$L$-radius $r_L$,
most probable initial radial coordinate $\hat{r}_0$ 
and number of trajectory loops $N_\mathrm{loops}$ 
for some values of the wave packet
parameters $p_0$, $\S$ and $j$.
Trajectories concerned in columns 6 to 9 are the most probable ones.}
\label{mean-values}
\end{center}\end{table}
Table \ref{mean-values} displays, for certain values of 
the wave parameters $p_0$, $\S$ and $j$, 
quantities of interest 
such as mean energy $\vev{E}$ \equ{meanE-explicit}, 
standard energy deviation $\D E$ \equ{DeltaE-explicit},
$\tau_\mathrm{min}$ \equ{time-incert},
(which are usual quantum theory quantities), and also, 
specifically concerning the most probable dBB trajectory, 
its approximate observed decay time $\tau_\mathrm{obs}$,
its $L$-radius $r_L$ \equ{L-radius}, 
its initial radial coordinate $\hat{r}_0$ (see \equ{r0hat})
which fixes it,
and the number $N_\mathrm{loops}$ of closed loops it performs before
passing to the straight-way regime. 

One can make the following  remarks about the items of Table
\ref{mean-values}. 
\begin{enumerate}
\item
The mean values $\vev{E}$, $\D E$, hence $\tau_\mathrm{min}$,
do not depend on the quantum number $j$, which is  obvious
from the explicit expressions \equ{meanE-explicit}  and 
\equ{DeltaE-explicit}. $\vev{E}$ and $\D E$ tend towards  
their limit values \equ{E-DE-approx} as the width $\S$ becomes
narrower, as can be seen in the Table.

\item The observed decay time $\tau_\mathrm{obs}$ seen in the behaviour of the 
azimuthal angle $\f_\mathrm{B}$ shown, \eg in
Figs. \ref{fig-gauss4c} or \ref{fig-gauss4d}, depends on the specific trajectory: it diminishes when the value of 
the initial radial coordinate
$r_\mathrm{B}(r_0,0)=r_0$ augments. 
On the other hand, its value does not depend sensibly on the
quantum number $j$, as can be seen in the table.

\item Except for $j=1/2$, the $L$-radius $r_L$ \equ{L-radius} is  
near of the value of the initial radial coordinate $\hat{r}_0$ of the most probable trajectory. This is what can be expected for the nearly circular motion which takes place at times $t<\tau_\mathrm{obs}$. 

\item  The number of revolutions 
 also tends to decrease with increasing initial position $r_0$, 
 as shown in the example of Fig. \ref{fig-gauss4}, which 
 shows five trajectories corresponding to five different 
 initial radial positions.

\item The behaviours observed in these examples
are generic, this being confirmed
by all other cases we have numerically studied.

\end{enumerate}

Concluding this subsection, an important observation can be made.
Although the minimum value for the decay-time $\tau$ was 
inferred from the usual quantum theoretical uncertainty principle 
for time-energy \equ{time-incert}, it appears difficult to interpret 
$\tau$  in this  
framework. But it looks quite natural in the dBB scheme, namely
as a property of the dBB trajectories.

\subsubsection{Times of flight}

The dBB theory also offers a very natural way to define the time of flight 
of a particle which has followed a dBB trajectory $\bx_\dbb(\bx_0,t)$ 
from its initial position $\bx_0$ to some target, 
\eg consisting of a detector. 
In our case, one can think of a detector occupying a 
circle of radius $R$ centered at the origin. This time of flight is 
then the solution $t_\mathrm{flight}(R,r_0)$ of the equation
\eq 
r_\mathrm{B}(r_0,t_\mathrm{flight})- R = 0,
\eqn{t-flight-eq}
where $r_\mathrm{B}(r_0,t)$ is the radial coordinate of the 
considered trajectory, characterized by its initial radial coordinate
$r_0$.
In case the solution is not unique, one has to take the lowest one, corresponding to the first hit of the particle to the 
target~\cite{Durr-etal,Das-etal}. However, this precaution is not needed in 
all cases we have investigated, where $r_\mathrm{B}(r_0,t)$ is 
a monotonically increasing function of $t$.

The empirically interesting quantity is the probability distribution 
$\Pi(\tau)$ in terms of the time of flight $\tfl=\tau$. 
It is given by the equation (9) of~\cite{Durr-etal}, which in our context 
takes the form:
\eq\ba{l}
\Pi(\tau)=N 2\pi\int_0^\infty dr_0\, 
r_0\, \rho(r_0,0)\,\d\lp \tfl(R,r_0)-\tau\rp\es
\phantom{\Pi(\tau)}=N 2\pi r_0(R,\tau)\,\rho(r_0(R,\tau),0)\,/\,
\left|\pa_{r_0}\tfl(R,r_0(R,\tau))\right|,
\ea\eqn{prob-t-flight}
where  $r_0(R,\tau)$ is the inverse of the time of flight function 
$\tfl(R,r_0)$, \ie the solution (unique, here) 
of \equ{t-flight-eq} for $r_0$ 
in terms of $R$ and $\tfl=\tau$. Recall that $\rho(r_0,0)$ 
represents the probability distribution for the trajectory defined 
by its initial radial coordinate $r_0$. 
$N$ is a normalization factor ensuring the normalization condition
\eq 
\int_0^\infty d\tau \, \Pi(\tau)=1.
\eqn{norm-Pi}
If the probability flux through the detector's entry
is always positive, which is the case in our 
examples\footnote{Examples where this is not the case 
are presented and discussed 
in~\cite{Durr-Teufel,Durr-etal,Das-etal,Sid-Das}.},
an alternative expression for the probability distribution is given 
by~\cite{Daumer-etal}
\eq 
\Pi_\mathrm{Flux}(\tau) = N \int_\S d\mathbf{s}\cdot 
\mathbf{j}(\bx,\tau)
\hspace{4mm} =^{^{\hspace{-7mm} 
\mbox{\small(here)}}} N 2\pi R j_r(R,\tau),
\eqn{Pi-flux}
where $\mathrm{j}$ is the probability flux and $\S$ 
the detector entry's surface. This result was proved in a scattering
context by~\cite{Daumer-etal}, and more generally, but in 
the one-dimensional case, by~\cite{Leavens}, and by~\cite{Das-Noeth}
in the case of a spinless non-relativistic particle. 
We have checked numerically the equivalence
of both formulae \equ{prob-t-flight} and \equ{Pi-flux}  in our specific situation for various parametrizations of the wave function. 
\begin{figure}[!h]
\begin{subfigure}[b]{.5\textwidth}
\centering
\includegraphics[width=.9\textwidth]{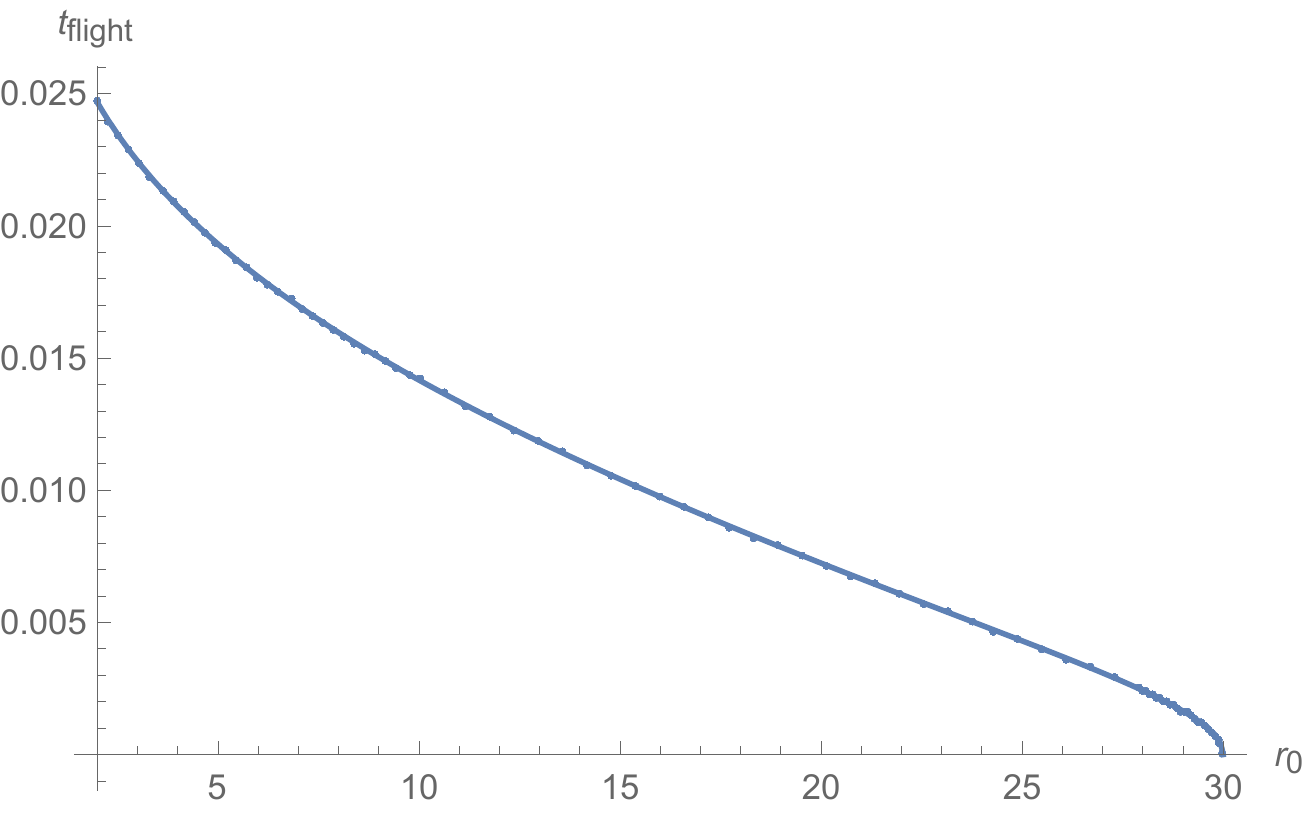}
\caption{} 
   \label{fig-gauss6a}
\end{subfigure}%
\begin{subfigure}[b]{.5\textwidth}
\centering
\includegraphics[width=1.\textwidth]{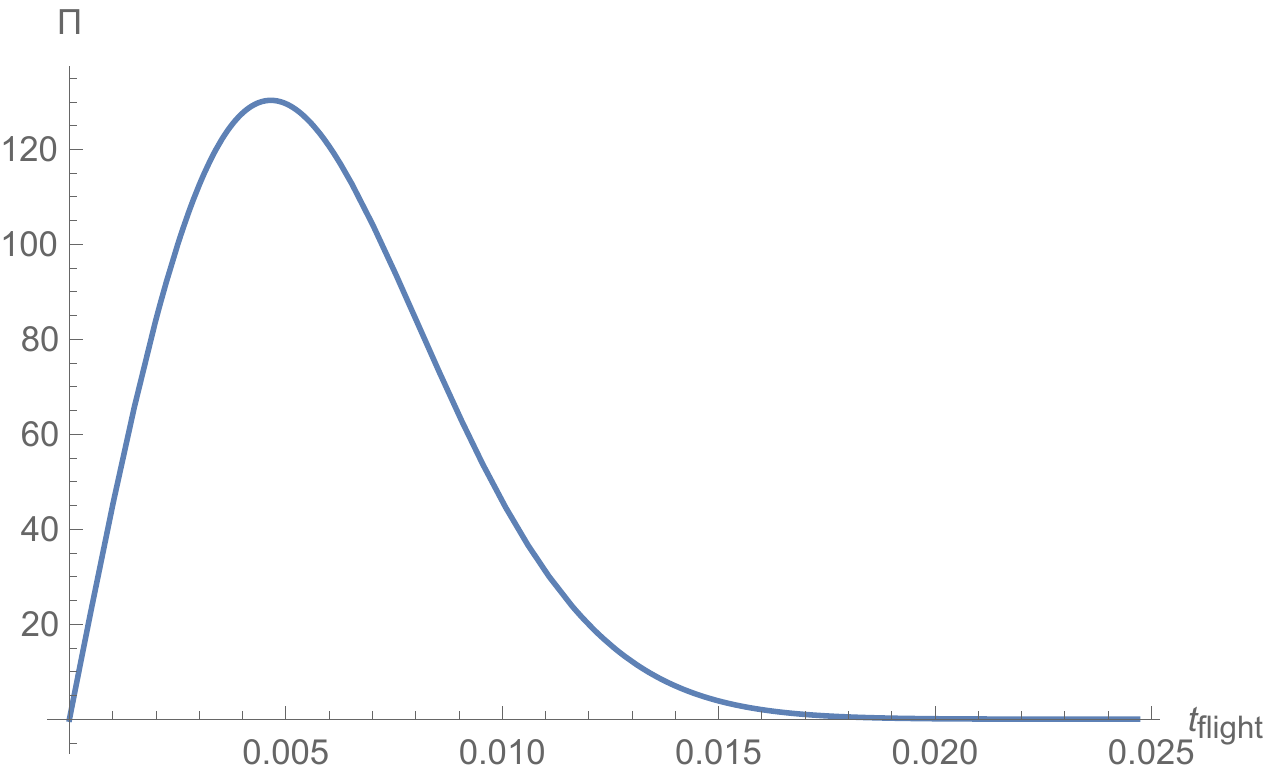}
\caption{} 
   \label{fig-gauss6b}
\end{subfigure}%
\caption{Times of flight  $t_\mathrm{flight}$ solutions of
\equ{t-flight-eq} and values of 
their probability density $\Pi(t_\mathrm{flight})$ \equ{prob-t-flight}
for the dBB trajectories shown in Fig. \ref{fig-gauss4}. 
The wave function parameters are the same as those in 
Figs. \ref{fig-gauss2} to \ref{fig-gauss5}. 
The target is a circle centred at the origin, with radius 
$R=30$ nm. The initial radial coordinate $r_0$ varies between 
2 and 30 nm.\\
(a) Values of the time of flight for
each dBB trajectory. The dots represent 
the numerically calculated values, and the continuous line an
interpolation used for the calculation of the probability 
distribution.\\
(b) Values of the corresponding probability density. 
Use of Eq. \equ{Pi-flux} has been made.}
\label{fig-gauss6}
\end{figure}
\begin{figure}[!h]
\begin{subfigure}[b]{.5\textwidth}
\centering
\includegraphics[width=.9\textwidth]{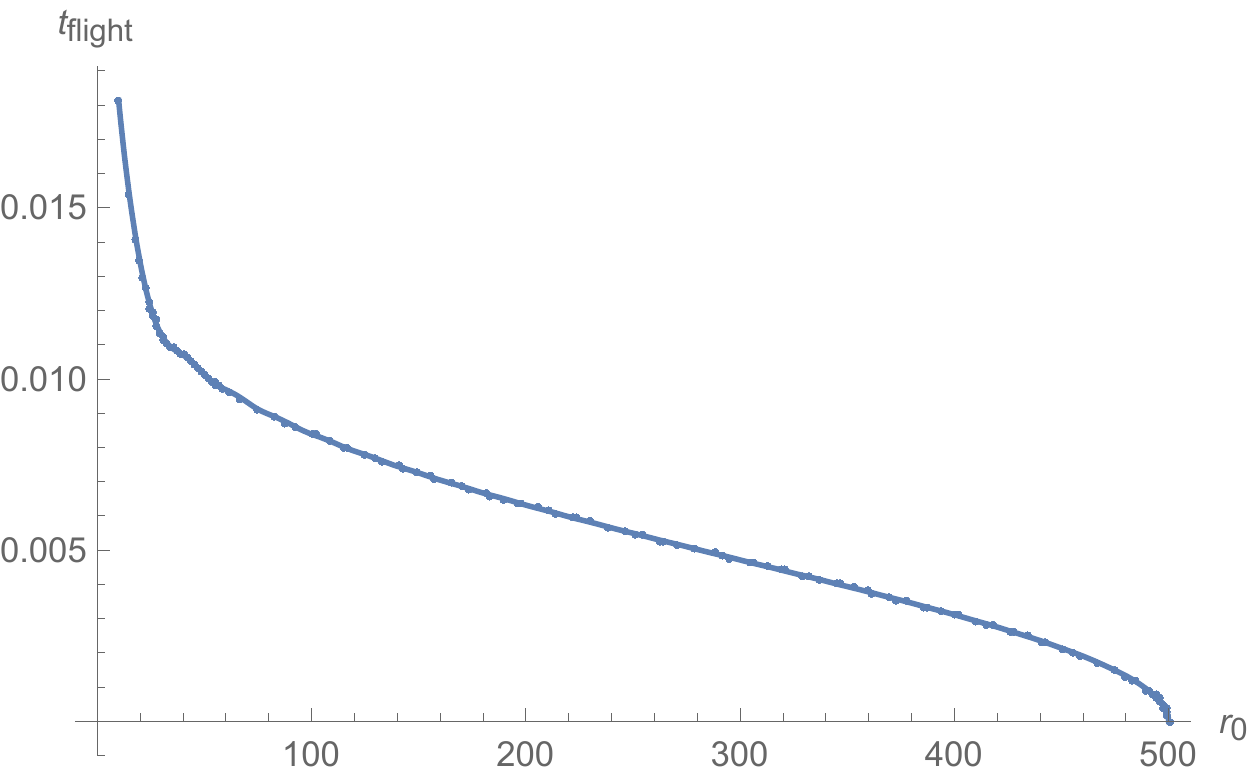}
\caption{} 
   \label{fig-gauss7a}
\end{subfigure}%
\begin{subfigure}[b]{.5\textwidth}
\centering
\includegraphics[width=1.\textwidth]{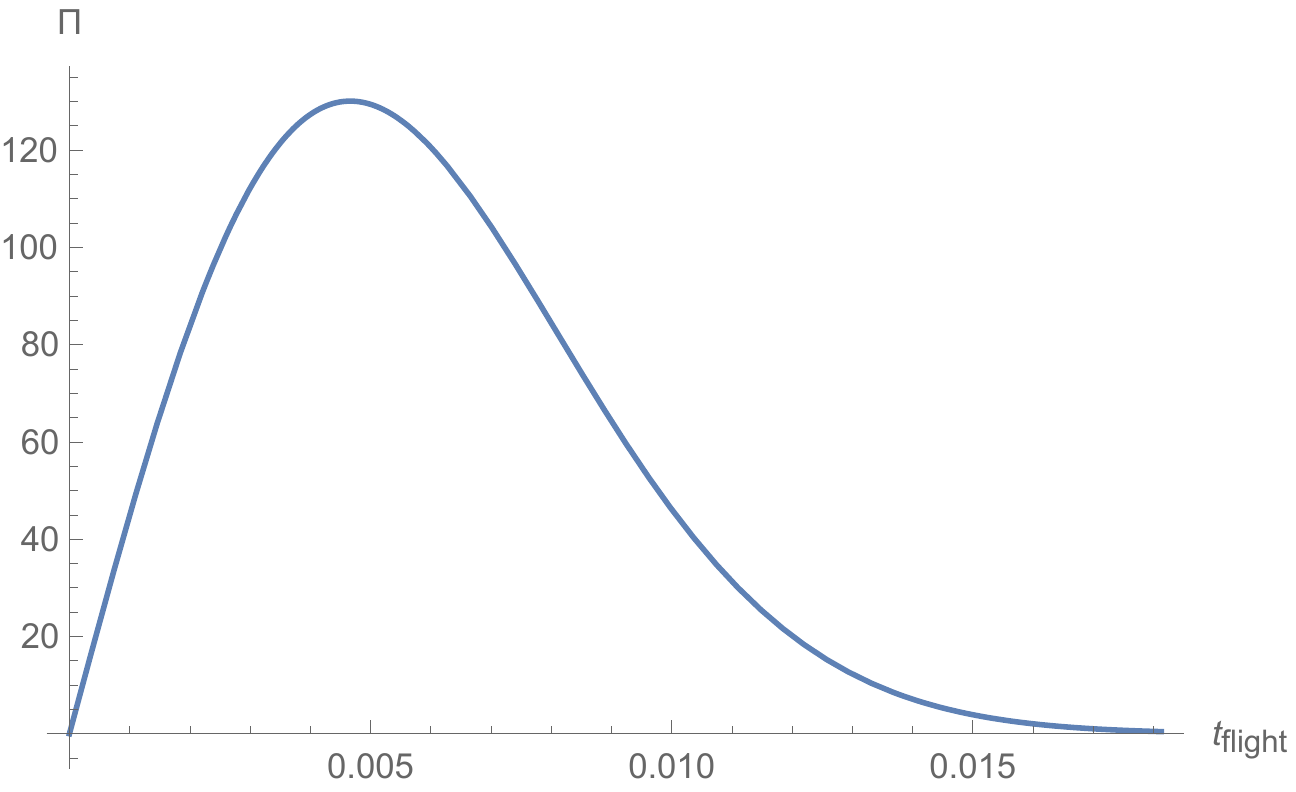}
\caption{} 
   \label{fig-gauss7b}
\end{subfigure}%
\caption{Same as Fig. \ref{fig-gauss6}, but 
with target's radius $R=500$ nm and initial radial coordinate 
$r_0$ in the interval 10 to 500 nm.\\
}
\label{fig-gauss7}
\end{figure}
Figs. \ref{fig-gauss6} and \ref{fig-gauss7} show the time of flight 
as a function of the
trajectory parameter $r_0$  and the corresponding 
probability distribution \equ{prob-t-flight} for
a circular target of radius 30 and 500 nm, respectively.

\section{Conclusion}

The trajectories predicted by the de Broglie-Bohm (dBB) quantum theory 
were calculated for the case of a guiding wave function 
being solution of the two-dimensional free Dirac equation, 
a solution constrained to be an eigenfunction of the total angular momentum operator
relative to a given origin point of space. 
Numerical results have been provided for the case of  massless particles with momentum-energy specifications corresponding to those of free electrons in mono-layer graphene.

The trajectories
corresponding to stationary wave functions turn out to be circles travelled at a constant speed. For Gaussian-like wave packets, the trajectories begin as quasi circles of slowly increasing radius till a critical time at which they tend to straight lines approximating the behaviour expected for a classical free particle. This transition time decreases when the value of the initial radial coordinate which labels a particular trajectory increases,
but appears to be insensible to the chosen value of the total angular momentum. It is worth noting that the transition time obtained in each example is of the order of magnitude of, but greater than, the lower bound given by the ''time-energy uncertainty principle''. Although the
nature of this lower bound is of course purely quantum mechanical, a theory such as the dBB one appears necessary in order to interpret it.
More, it is the use of the dBB theory which has allowed us to put this phenomenon in evidence.

Given a wave function, the possible times of arrival of the particle at some region have also been calculated as a function of its initial position 
for the same examples, taking profit of the objective reality of the trajectories in the dBB theory. The corresponding probability distribution of these arrival times has been calculated  using the Das-D\"urr formula \equ{prob-t-flight} and also  the probability flux
formula \equ{Pi-flux}. Both calculation's results coincide, as can be expected from the equivalence's proof given 
in~\cite{Leavens} for the spin one-half particle in 
one-dimensional space and by~\cite{Das-Noeth} for the non-relativistic spinless particle. Note that this equivalence holds if the flux on any target is always positive -- which is true in our examples. The interest of this probability distribution is that it may in principle be measured in a given 
physical context such as, \eg the mono-layer graphene, after a suitable analysis of the experimental setting.

\subsubsection*{Acknowledgements}
I would like to thank Siddhant Das for his reading of the manuscript,
the indication of 
interesting references and  valuable comments.


\appendix
\section*{Appendices}

\section{Notations and conventions}\label{not-conv}

Units used in this paper are adapted to the physics of graphene. Length, 
time and energy are given in nm, ns and meV, respectively. 
The critical velocity and the Planck constant take the values
\eq
c=10^6\,\mbox{nm \,ns}^{-1},
\quad \hbar=6.5821 \times 10^{-4}\mbox{ meV ns}.
\eqn{fund-const-graf}
Space-time coordinate are denoted by $x^\m$, $\m=0,1,2$, 
space coordinates by $\bx=(x,y)$, or $(r,\f)$. Space-time metric is
$\eta_{\m\n}$ = diag$(1,-1,-1)$

Dirac matrices are chosen in terms of the Pauli matrices as
\eq
\g^0=\s^z,\quad \g^1=\g^0\s^x ,\quad \g^2=\g^0\s^y.
\eqn{gamma-mat} 
The Dirac matrices  $\a^i=\g^0\g^i$ used in the non-relativistic formulation explicitly are
\eq
\a^1=\s^x,\quad \a^2=\s^y,\quad \b=\s^z.
\eqn{alpha-mat}

\section{Some useful properties of the Bessel functions}
\label{app-bessel}

The general solution of the Bessel equation~\cite{Bessel} 
\eq 
z^2 f''(z)+z f'(z) + (z^2-n^2)f(z) = 0,
\eqn{bessel-ew}
has the form
\eq 
f(z)=C_1 J_n(z)+C_2 Y_n(z),
\eqn{bessel-gen-sol}
where $J_n$ and $Y_n$ are the Bessel functions of the 
first~\cite{Bessel}, respectively second~\cite{Bessel} kind, 
and $C_1$, $C_2$ are two arbitrary complex constants.
We shall  restrict ourselves to an integer index $n$.

The asymptotic behaviours of the Bessel functions at the origin  are given by
\eq\ba{ll} 
J_n(x)\sim\dfrac{1}{n!}\lp\dfrac{x}{2}\rp^n   
\quad &(0<x\ll1,\ n\ge0),\es
Y_n(x)\sim -\dfrac{(n-1)!}{\pi}\lp\dfrac{2}{x}\rp^n 
\quad &(0<x\ll1,\ n\ge1),\es
Y_0(x)\sim \dfrac{2}{\pi} \log\lp\dfrac{x}{2}\rp     \quad & (0<x\ll1),
\ea\eqn{as-origin} 
and at infinity by
\eq\ba{ll} 
J_n(x)\sim\sqrt{\dfrac{2}{\pi x}}\cos\lp x-\dfrac{(n+\frac12)\pi}{2}\rp    
\qquad &(x\gg1,\ n\ge0) ,\es
 Y_n(x)\sim \sqrt{\dfrac{2}{\pi x}}\sin\lp x-\dfrac{(n+\frac12)\pi}{2}\rp       
 \qquad &(x\gg1,\ n\ge0).
\ea\eqn{as-infinity} 
Functions with a negative index  are related to those with a positive  one by the identities
\eq 
J_{-n}(z)=(-1)^n J_n(z),\qquad Y_{-n}(z)=(-1)^n Y_n(z).
\eqn{pos<->neg-index}
Under parity $z\to-z$, the function $J_n$ transforms as
\eq
J_n(-z)=(-1)^n J_n(z).
\eqn{BesselJ-parity}
An interesting orthogonality property is given by~\cite{Bessel}
\eq 
\dint_{\!\!\!\!0}^R dr\,r\,
J_n\lp\frac{z_{n,\a}\,r}{R}\rp J_n\lp\frac{z_{n,\b}\,r}{R}\rp 
=\frac{R^2}{2}\lp J_{n+1}(z_{n,\a}\rp^2 \d_{\a\b},
\eqn{Bessel-ortho}
for $n\ge0$, where $z_{n,\a}$ is the $\a^{\rm th}$ positive zero 
of the Bessel function $J_n(z)$~\cite{Bessel-zeros}. 
Moreover, any function
$f(r)$ defined in the interval $0\le r\le R$ with bounded variation 
and vanishing at the end point $r=R$ can be represented as a ``Fourier Bessel series''~\cite{Weisstein} as
\eq 
f(r)=\sum_{\a=1}^\infty c_\a J_n\lp\frac{z_{n,\a}\,r}{R}\rp,
\eqn{Fourier-Bessel}
for any $n\ge0$. The coefficients $c_\a$ can be calculated using the orthogonality formula \equ{Bessel-ortho}.

\section*{Declarations}

No funds, grants, or other support was received.

\noindent The author has no relevant financial or non-financial interests to disclose.

\noindent The Mathematica notebook file relevant  for the 
numerical calculations made in this research may be required 
at author's e-mail: {\em opiguet@yahoo.com}.



\begin{thebibliography}{999}

\bibitem{Planck} Max Planck, ``Ueber das Gesetz der Energieverteilung im Normalspectrum'' (English translation), {\em Annalen der Physik} 4 (1901) 553.

\bibitem{Bohr} Niels Bohr, ``On the Constitution of Atoms and Molecules'', {\em Philos. Mag.} 26 (1913) 1  and  476.

\bibitem{Einstein} Albert Einstein, ``Concerning an Heuristic Point of View Toward the Emission and Transformation of Light'', {\em Annalen der Physik} 17 (1905) 132.

\bibitem{deBroglie} Louis de Broglie, ``Recherches sur la th\'eorie des quanta'', Thesis (Paris), 1924;\\ 
Louis de Broglie, {\em Ann. Phys. (Paris)} 3, 22 (1925). 
Reprint in {\em Ann. Found. Louis de Broglie} 17 (1992) p. 22;\\
Louis De Broglie, ``La m\'ecanique ondulatoire et la structure 
atomique de la mati\`ere et du rayonnement'', 
{\em J. Phys. Radium} 8 (1927) 225, DOI 10.1051/jphysrad:0192700805022500.

\bibitem{Schroedinger}Erwin Schr\"odinger, ``Quantisierung als Eigenwertproblem'',  {\em  Annalen der Physik} 79 (1926), 361,
{\em  Annalen der Physik} 79 (1926) 489,
{\em  Annalen der Physik} 80 (1926) 437,
  {\em  Annalen der Physik} 81 (1926) 109.

\bibitem{Heisenberg} Werner Heisenberg, ``Über quantentheoretische Umdeutung kinematischer und mechanischer Beziehungen'', \zp{33}{1925}{879}.

\bibitem{Dirac} Paul A.M. Dirac, ``The quantum theory of the electron'',
{\em Proc. R. Soc.} A 117 (1928) 610 and  118 (1928) 351.

\bibitem{Copenhagen-int} Niels Bohr, ``The Quantum Postulate and the Recent Development of Atomic Theory'', {\em Supplement to "Nature} April 14 (1928) 580;\\
Werner Heisenberg, ``Physics and Philosophy'', Harper, New York (1958), 

\bibitem{Manyworld-int} Hugh Everett, ``Relative State Formulation of Quantum Mechanics'', \\  \rmp{29}{1957}{454}.

\bibitem{Relational} Carlo Rovelli, ``Relational quantum mechanics'',\\
\ijtp{35}{1996}{1637} e-Print: quant-ph/9609002 [quant-ph];\\
Andrea Di Biagio and Carlo Rovelli, ``Stable Facts, Relative Facts'', \\
\Foundation{}{2021}51:30.

\bibitem{Bohm} David Bohm, ``A Suggested interpretation of the quantum theory in terms of hidden variables 1, 2.'', \pr{85}{1952}{166,\,180}.

\bibitem{Bohm-Hiley} D. Bohm and B.J. Hiley, "The Undivided Universe", Routledge, London and New York (1995).

\bibitem{Bell-book} John S. Bell, 
``Speakable and Unspeakable in Quantum Mechanics'', 
{\em Cambridge University Press, New York} (2010). 

\bibitem{Daumer-etal} M. Daumer, D. D\"urr, S. Goldstein and 
N. Zanghi, ``On the Quantum Probability Flux Through Surfaces'',
\jsp{88}{1997}{967}.

\bibitem{Durr-etal} Siddhant Das and Detlef D\"urr, 
"Arrival time distributions 
of spin-1/2 particles, {\em Nature Scient. Rep.} 9\,(2019)\, 2242.

\bibitem{Das-etal} Siddhant Das, Markus Nöth and Detlef D\"urr,
``Exotic arrival times of spin-1/2 particles I - 
An analytical treatment'', \pr{A99}{2019}{052124}.

\bibitem{Holland-book} Peter R. Holland, 
``The quantum theory of motion'', Revised ed.,  
Cambridge University Press  (1995).

\bibitem{Durr-etal-relativity} Detlef D\"urr, Sheldon Goldstein, 
Travis Norsen, Ward Struyve and Nino Zanghì,
``Can Bohmian mechanics be made relativistic?''
{\em Proc. R. Soc.} A 470\,(2013)\,20130699

\bibitem{Tumulka} Roderich Tumulka, ``On Bohmian Mechanics, 
Particle Creation, and Relativistic Space-Time: Happy 100th 
Birthday, David Bohm!'', 
{\em Entropy} 20\,(2018)\,462.

\bibitem{Bressanini-Ponti} Dario Bressanini and Alessandro Ponti,
``Angular Momentum and the Two-Dimensional Free Particle´´,
{\em J. Chem. Educ.} 75\,(1998)\,916.

\bibitem{Mathematica} Wolfram Research, 
Inc., Mathematica, Champaign, IL.
 
\bibitem{APais} Abraham Pais, ``On Spinors in n Dimensions'',
\jmp{3}{1962}{1135}; doi: 10.1063/1.1703856.

\bibitem{Durr-Goldstein-Zanghi} D. D\"urr D., S. Goldstein  and N. Zanghi,
 ``Quantum equilibrium and the origin of absolute uncertainty", 
\jsp{67}{1992}{843}.

\bibitem{Holland} P.R. Holland,
``The Dirac equation in the de Broglie-Bohm theory of motion'',
\Foundation{22}{1992}{1287}.

\bibitem{compl-Bessel} The Wolfram Functions Site,\\ 
\url{https://functions.wolfram.com/Bessel-TypeFunctions/BesselJ/21/02/02/}

\bibitem{error-function} WolframMathWorld,\\
\url{https://functions.wolfram.com/GammaBetaErf/Erf/}

\bibitem{Katsnelson} Mikhail I. Katsnelson, 
``The Physics of Graphene'' 2nd Edition, Cambridge University Press, Cambridge (2020).

\bibitem{Das-Sarma} S. Das Sarma, Shaffique Adam, E. H. Hwang
and Enrico Rossi, ``Electronic transport in two-dimensional graphene'',
\rmp{83}{2011}{407}, e-Print: arXiv:1003.4731 (cond-mat.mtrl-sci).

\bibitem{time-energy-uncertainty} Albert Messiah, ``Quantum Mechanics'',
Vol. 1, Section VIII-13, Dover Publications, New York (2014) (English translation 
of  ``M\'ecanique Quantique'', Dunod, Paris, (1962)).

\bibitem{Durr-Teufel} Detlef D\"urr and Stefan Teufel,
`Bohmian Mechanics'', Chap 16, Springer, Heidelberg (2009).

\bibitem{Sid-Das} Siddhant Das, ``Relativistic electron 
wave packets featuring quantum backflow'', {\em e-print}
arXiv:2112.13180.

\bibitem{Leavens} C. Richard Leavens, 
``Bohm Trajectory Approach to Timing Electrons'', 
p. 129 of ``Time in Quantum Mechanics - Vol.1'', 2$^{d}$ Ed., J.G. Muga,
R. Sala Mayato, \'I.L. Egusquiza (Eds.), 
{\em Lecture Notes in Physics} 734, Springer, Heidelberg, 2008.

\bibitem{Das-Noeth} Siddhant Das  and Markus Nöth, 
Times of arrival and gauge invariance, 
{\em Proc. R. Soc.} A 477\,(2021)\,20210101, 
https://doi.org/10.1098/rspa.2021.0101.

\bibitem{Bessel} WolframMathWorld, ``Bessel function'',\\
\url{https://mathworld.wolfram.com/topics/BesselFunctions.html}.
 
\bibitem{Bessel-zeros} WolframMathWorld,\\
\url{https://mathworld.wolfram.com/BesselFunctionZeros.html}
 
\bibitem{Weisstein} Eric W. Weisstein,  ``Fourier-Bessel Series.'' From MathWorld--A Wolfram Web Resource. https://mathworld.wolfram.com/Fourier-BesselSeries.html 


\end{thebibliography}
\end{document}